\documentclass[aps,eqsecnum,preprint,floats,epsf,epsfig,nofootinbib,letter]{revtex4}
\usepackage{epsfig}

\begin{document}
\def\be{\begin{eqnarray}}
\def\en{\end{eqnarray}}
\def\non{\nonumber}
\def\la{\langle}
\def\ra{\rangle}
\def\nc{N_c^{\rm eff}}
\def\vp{\varepsilon}
\def\drho{\bar\rho}
\def\deta{\bar\eta}
\def\CP{{\it CP}~}
\def\A{{\cal A}}
\def\B{{\cal B}}
\def\C{{\cal C}}
\def\d{{\cal D}}
\def\e{{\cal E}}
\def\p{{\cal P}}
\def\t{{\cal T}}
\def\up{\uparrow}
\def\dw{\downarrow}
\def\vma{{_{V-A}}}
\def\vpa{{_{V+A}}}
\def\smp{{_{S-P}}}
\def\spp{{_{S+P}}}
\def\J{{J/\psi}}
\def\ov{\overline}
\def\Lqcd{{\Lambda_{\rm QCD}}}
\def\pr{{Phys. Rev.}~}
\def\prl{{Phys. Rev. Lett.}~}
\def\pl{{Phys. Lett.}~}
\def\np{{Nucl. Phys.}~}
\def\zp{{Z. Phys.}~}
\def\lsim{ {\ \lower-1.2pt\vbox{\hbox{\rlap{$<$}\lower5pt\vbox{\hbox{$\sim$}
}}}\ } }
\def\gsim{ {\ \lower-1.2pt\vbox{\hbox{\rlap{$>$}\lower5pt\vbox{\hbox{$\sim$}
}}}\ } }

\font\el=cmbx10 scaled \magstep2{\obeylines\hfill October, 2007}

\vskip 1.5 cm

\centerline{\large\bf Hadronic charmless $B$ decays $B\to AP$}
\bigskip
\centerline{\bf Hai-Yang Cheng$^{1}$ and Kwei-Chou Yang$^{2}$}
\medskip
\centerline{$^1$ Institute of Physics, Academia Sinica}
\centerline{Taipei, Taiwan 115, Republic of China}
\medskip
\centerline{$^2$ Department of Physics, Chung Yuan Christian
University} \centerline{Chung-Li, Taiwan 320, Republic of China}
\bigskip
\bigskip
\bigskip
\bigskip
\centerline{\bf Abstract}
\bigskip
\small
 The two-body hadronic decays of $B$ mesons into pseudoscalar and
axial vector mesons are studied within the framework of QCD
factorization. The light-cone distribution amplitudes (LCDAs) for
$^3P_1$ and $^1P_1$ axial-vector mesons have been evaluated using
the QCD sum rule method. Owing to the $G$-parity, the chiral-even
two-parton light-cone distribution amplitudes of the $^3P_1$
($^1P_1$) mesons are symmetric (antisymmetric) under the exchange of
quark and anti-quark momentum fractions in the SU(3) limit. For
chiral-odd LCDAs, it is other way around. The main results are the
following: (i) The predicted rates for $a_1^\pm(1260)\pi^\mp$,
$b_1^\pm(1235)\pi^\mp$, $b_1^0(1235)\pi^-$, $a_1^+K^-$ and
$b_1^+K^-$ modes are in good agreement with the data. However, the
naively expected ratios $\B(B^-\to a_1^0\pi^-)/\B(\ov B^0\to
a_1^+\pi^-)\lsim 1$, $\B(B^-\to a_1^-\pi^0)/\B(\ov B^0\to
a_1^-\pi^+)\sim {1\over 2}$ and $\B(B^-\to b_1^0K^-)/\B(\ov B^0\to
b_1^+K^-)\sim {1\over 2}$ are not borne out by experiment. This
should be clarified by the improved measurements of these decays.
(ii) Since the $\ov B\to b_1K$ decays receive sizable annihilation
contributions, their rates are sensitive to the interference between
penguin and annihilation terms. The measurement of $\B(\ov B^0\to
b_1^+K^-)$ implies a destructive interference which in turn
indicates that the form factors for $B\to b_1$ and $B\to a_1$
transitions are of opposite signs. (iii) Sizable power corrections
such as weak annihilation are needed to account for the observed
rates of the penguin-dominated modes $K_1^-(1270)\pi^+$ and
$K_1^-(1400)\pi^+$. (iv) The decays $B\to K_1\ov K$ with
$K_1=K_1(1270),K_1(1400)$ are in general quite suppressed, of order
$10^{-7}\sim 10^{-8}$, except for $\ov B^0\to \ov K_1^0(1270)K^0$
which can have a branching ratio of order $2.3\times 10^{-6}$. The
decay modes $K_1^-K^+$ and $K_1^+K^-$ are of particular interest as
they proceed only through weak annihilation. (v) The mixing-induced
parameter $S$ is predicted to be negative in the decays $B^0\to
a_1^\pm\pi^\mp$, while it is positive experimentally. This may call
for a larger unitarity angle $\gamma\gsim 80^\circ$. (vi) Branching
ratios for the decays $B\to f_1\pi$, $f_1K$, $h_1\pi$ and $h_1K$
with $f_1=f_1(1285),f_1(1420)$ and $h_1=h_1(1170),h_1(1380)$  are
generally of order $10^{-6}$ except for the color-suppressed modes
$f_1\pi^0$ and $h_1\pi^0$ which are suppressed by one to two orders
of magnitude. Measurements of the ratios $\B(B^-\to h_1(1380)
\pi^-)/\B(B^-\to h_1(1170)\pi^-)$ and $\B(\ov B\to f_1(1420)\ov
K)/\B(\ov B\to f_1(1285)\ov K)$ will help determine the mixing
angles $\theta_{^1P_1}$ and $\theta_{^3P_1}$, respectively.

\pagebreak

\section{Introduction}
In the quark model, two nonets of $J^P=1^+$ axial-vector mesons
are expected as the orbital excitation of the $q\bar q$ system. In
terms of the spectroscopic notation $^{2S+1}L_J$, there are two
types of $p$-wave mesons, namely, $^3P_1$ and $^1P_1$. These two
nonets have distinctive $C$ quantum numbers, $C=+$ and $C=-$,
respectively. Experimentally, the $J^{PC}=1^{++}$ nonet consists
of $a_1(1260)$, $f_1(1285)$, $f_1(1420)$ and $K_{1A}$, while the
$1^{+-}$ nonet has $b_1(1235)$, $h_1(1170)$, $h_1(1380)$ and
$K_{1B}$. The physical mass eigenstates $K_1(1270)$ and
$K_1(1400)$ are a mixture of $K_{1A}$ and $K_{1B}$ states owing to
the mass difference of the strange and non-strange light quarks.

The production of the axial-vector mesons has been seen in the
two-body hadronic $D$ decays: $D\to Ka_1(1260)$, $D^0\to
K_1^-(1270)\pi$ and $D^+\to K_1^0(1400)\pi$ , and in charmful $B$
decays: $B\to J/\psi K_1(1270)$ and $B\to Da_1(1260)$ \cite{PDG}.
As for charmless hadronic $B$ decays, $B^0\to
a_1^\pm(1260)\pi^\mp$ are the first modes measured by both $B$
factories, BaBar and Belle. The BaBar result is \cite{BaBara1pi}
 \be
 \B(B^0\to a_1^\pm(1260)\pi^\mp)=(33.2\pm3.8\pm3.0)\times 10^{-6}.
 \en
where the assumption of $\B(a_1^\pm\to \pi^\pm\pi^\pm\pi^\mp)=1/2$
has been made. The Belle measurement gives \cite{Bellea1pi}
 \be
 \B(B^0\to a_1^\pm(1260)\pi^\mp)=(29.8\pm3.2\pm4.6)\times 10^{-6}.
 \en
The average of the two experiments is
 \be
 \B(B^0\to a_1^\pm(1260)\pi^\mp)=(31.7\pm3.7)\times 10^{-6}.
 \en
Moreover, BaBar has also measured the time-dependent $CP$
asymmetries in $B^0\to a_1^\pm(1260)\pi^\mp$ decays
\cite{BaBara1pitime}. From the measured \CP parameters, one can
determine the decay rates of $a_1^+\pi^-$ and $a_1^-\pi^+$
separately \cite{BaBara1pitime}. Recently, BaBar has reported the
observation of the decays $\ov B^0\to b_1^\pm\pi^\mp,b_1^+K^-$ and
$B^-\to b_1^0\pi^-,b_1^0K^-,a_1^0\pi^-,a_1^-\pi^0$
\cite{BaBarb1pi,BaBara1pi07}. The preliminary BaBar results for $\ov
B^0\to K_1^-(1270)\pi^+,K_1^-(1400)\pi^+,a_1^+K^-$, $B^-\to a_1^-\ov
K^0,f_1(1285)K^-,f_1(1420)K^-$ are also available recently
\cite{BaBara1K,Blanc,Brown,Burke}.

In the present work we will focus on the $B$ decays involving an
axial-vector meson $A$ and a pseudoscalar meson $P$ in the final
state.  Since the $^3P_1$ meson behaves similarly to the vector
meson, it is naively expected that $AP$ modes have similar rates as
$VP$ ones, for example, $\B(B^0 \to a_1^\pm(1260)\pi^\mp)\sim
\B(B^0 \to \rho^\pm\pi^\mp)$. However, this will not be the case
for the $^1P_1$ meson. First of all, its decay constant vanishes in
the SU(3) limit. For example, the decay constant vanishes for the
neutral $b_1^0(1235)$ and is very small for the charged $b_1(1235)$
states. This feature can be checked experimentally by measuring
$B^0\to b_1^+\pi^-,b_1^-\pi^+$ decays and seeing if the former is
suppressed relative to the latter. Second, its chiral-even
two-parton light-cone distribution amplitude (LCDA) is
anti-symmetric under the exchange of quark and anti-quark momentum
fractions in the SU(3) limit due to the $G$ parity, contrary to the
symmetric behavior for the $^3P_1$ meson.

Charmless $B\to AP$ and $B\to AV$ decays have been studied in the
literature
\cite{Yang1P1,Chen05,Nardulli05,Nardulli07,Calderon,Yanga1}.
Except for \cite{Yang1P1,Yanga1}, most of the existing
calculations were carried out in the framework of either naive
factorization or generalized factorization in which the
nonfactorizable effects are described by the parameter $N_c^{\rm
eff}$, the effective number of colors. In the approach of QCD
factorization, nonfactorizable effects such as vertex corrections,
hard spectator interactions and annihilation contributions are
calculable and have been considered in \cite{Yang1P1,Yanga1} for
the decays $B\to a_1(1260)\pi,~a_1(1260)K$, $B\to
h_1(1235)K^*,~b_1(1235)K^*$ and $b_1(1235)\rho$.

One crucial ingredient in QCDF calculations is the LCDAs for
$^3P_1$ and $^1P_1$ axial-vector mesons. In general, the LCDAs are
expressed in terms of the expansion of Gegenbauer moments which
have been systematically studied by one of us (K.C.Y.) using the
light-cone sum rule method \cite{YangLCDA1P1,YangNP}. Armed with
the LCDAs, one is able to explore the nonfactorizable corrections
to the naive factorization.

The present paper is organized as follows. In Sec. II we summarize
all the input parameters relevant to the present work, such as the
mixing angles, decay constants, form factors and light-cone
distribution amplitudes for $^3P_1$ and $^1P_1$ axial-vector
mesons. We then apply QCD factorization in Sec. III to study $B\to
AP$ decays. Results and discussions are presented in Sec. IV. Sec.
V contains our conclusions. The factorizable amplitudes of various
$B\to AV$ decays are summarized in Appendix A.

\section{Input parameters}

\subsection{Mixing angles}

In the quark model, there are two different types of light axial
vector mesons: $^3P_1$ and $^1P_1$, which carry the quantum numbers
$J^{\rm PC}=1^{++}$ and $1^{+-}$, respectively. The $1^{++}$ nonet
consists of $a_1(1260)$, $f_1(1285)$, $f_1(1420)$ and $K_{1A}$,
while the $1^{+-}$ nonet has $b_1(1235)$, $h_1(1170)$, $h_1(1380)$
and $K_{1B}$. The non-strange axial vector mesons, for example, the
neutral $a_1(1260)$ and $b_1(1235)$ cannot have mixing because of
the opposite $C$-parities. On the contrary, the strange partners of
$a_1(1260)$ and $b_1(1235)$, namely, $K_{1A}$ and $K_{1B}$,
respectively, are not mass eigenstates and they are mixed together
due to the strange and non-strange light quark mass difference. We
write
 \be \label{mixing}
 K_1(1270) &=& K_{1A}\sin\theta_{K_1}+K_{1B}\cos\theta_{K_1}, \non \\
 K_1(1400) &=& K_{1A}\cos\theta_{K_1}-K_{1B}\sin\theta_{K_1}.
 \en
If the mixing angle is $45^\circ$ and $\la K\rho|K_{1B}\ra=\la
K\rho|K_{1A}\ra$, one can show that $K_1(1270)$ is allowed to
decay into $K\rho$ but not $K^*\pi$, and vice versa for
$K_1(1400)$ \cite{Lipkin77}.

From the experimental information on masses and the partial rates of
$K_1(1270)$ and $K_1(1400)$, Suzuki found two possible solutions
with a two-fold ambiguity, $|\theta_{K_1}|\approx 33^\circ$ and
$57^\circ$ \cite{Suzuki}. A similar constraint $35^\circ\lsim
|\theta_{K_1}|\lsim 55^\circ$ is obtained in \cite{Goldman} based
solely on two parameters: the mass difference of the $a_1$ and $b_1$
mesons and the ratio of the constituent quark masses. From the data
of $\tau\to K_1(1270)\nu_\tau$ and $K_1(1400)\nu_\tau$ decays, the
mixing angle is extracted to be $\pm37^\circ$ and $\pm 58^\circ$ in
\cite{ChengDAP}. As for the sign of the mixing angle, there is an
argument favoring a negative $\theta_{K_1}$. It has been pointed out
in \cite{ChengKgamma} that the experimental measurement of the ratio
of $K_1\gamma$ production in $B$ decays can be used to fix the sign
of the mixing angle. Based on the covariant light-front quark model
\cite{CCH}, it is found \cite{ChengKgamma} \footnote{The sign of
$\theta_{K_1}$ is intimately related to the relative sign of the
$K_{1A}$ and $K_{1B}$ states. In the light-front quark model used in
\cite{ChengKgamma} and \cite{CCH}, the decay constants of $K_{1A}$
and $K_{1B}$ are of opposite sign, while the $B\to K_{1A}$ and $B\to
K_{1B}$ form factors are of the same sign. It is other way around in
the present work: the decay constants of $K_{1A}$ and $K_{1B}$ have
the same signs, while the $B\to K_{1A}$ and $B\to K_{1B}$ form
factors are opposite in sign. The two schemes are related via a
redefinition of the $K_{1A}$ or $K_{1B}$ state, i.e. $K_{1A}\to
-K_{1A}$ or $K_{1B}\to -K_{1B}$. To write down
Eq.(\ref{eq:RK1gamma}) we have used our convention for $K_{1A}$ and
$K_{1B}$ states.}
 \be \label{eq:RK1gamma}
 {\B(B\to K_1(1270)\gamma)\over \B(B\to K_1(1400)\gamma)}=\cases{
 10.1\pm6.2~~(280\pm200); & for~$\theta_{K_1}=-58^\circ~(-37^\circ)$, \cr
 0.02\pm0.02~(0.05\pm0.04); & for~$\theta_{K_1}=+58^\circ~(+37^\circ)$.}
 \en
The Belle measurements $\B(B^+\to
K_1^+(1270)\gamma)=(4.3\pm0.9\pm0.9)\times 10^{-5}$ and $\B(B^+\to
K_1^+(1400)\gamma)<1.5\times 10^{-5}$ \cite{BelleK1gamma} clearly
favor $\theta_{K_1}=-58^\circ$ over $\theta_{K_1}=58^\circ$ and
$\theta_{K_1}=-37^\circ$ over $\theta_{K_1}=37^\circ$.
In the ensuing
discussions we will fix the sign of $\theta_{K_1}$ to be negative.

Likewise, the $^3P_1$ states $f_1(1285)$ and $f_1(1420)$ have
mixing due to SU(3) breaking effects
\begin{eqnarray}
 |f_1(1285)\rangle = |f_1\rangle\cos\theta_{^3P_1}+|f_8\rangle\sin\theta_{^3P_1},
 \quad |f_1(1420)\rangle =
 -|f_1\rangle\sin\theta_{^3P_1} +|f_8\rangle\cos\theta_{^3P_1} \,.
 \end{eqnarray}
From the Gell-Mann-Okubo mass formula \cite{CloseBook,PDG}, it
follows that
\begin{equation}
\cos^2\theta_{^3P_1}
=\frac{4m_{K_{1A}}^2-m_{a_1}^2-3m_{f_1(1285)}^2}
{3\bigg(m_{f_1(1420)}^2-m_{f_1(1285)}^2\bigg)}\,,\label{eq:GMOkubo}
\end{equation}
where
 \begin{equation}
 m_{K_{1A}}^2 = m_{K_1(1400)}^2 \cos^2\theta_{K_1} + m_{K_1(1270)}^2
 \sin^2\theta_{K_1} \,.
 \end{equation}
Substituting this into Eq.~(\ref{eq:GMOkubo}) with $\theta_{K_1}
=-37^\circ\,(-58^\circ)$, we then obtain $\theta_{^3P_1}^{\rm quad}=
27.9^\circ\,(53.2^\circ)$ and $\theta_{^3P_1}^{\rm lin}=
26.0^\circ\,(52.1)^\circ$ where the latter is obtained by replacing
the meson mass squared $m^2$ by $m$ throughout
Eq.~(\ref{eq:GMOkubo}). The sign of the mixing angle can be
determined from the mass relation \cite{PDG}
\begin{equation} \label{eq:GMOkubolinear}
 \tan\theta_{^3P_1}
=\frac{4m_{K_{1A}}^2-m_{a_1}^2-3m_{f_1(1420)}^2}
{2\sqrt{2}(m_{a_1}^2-m_{K_{1A}}^2)}\,.
\end{equation}
The previous phenomenological analyses suggest that $\theta_{^3P_1}
\simeq 50^\circ$ \cite{Close:1997nm}. \footnote{If a mixing angle
$\theta_{^3P_1}$  of order $50^\circ$ can be independently inferred
from other processes, this will imply a preference of
$|\theta_{K_1}|=58^\circ$ over $|\theta_{K_1}|=37^\circ$. However,
the phenomenological analysis in \cite{Close:1997nm} is not robust
and the Gell-Mann-Okubo mass formula employed there is not a correct
one.}
Eliminating $\theta$ from Eqs. (\ref{eq:GMOkubo}) and
(\ref{eq:GMOkubolinear}) leads to the sum rule
 \be
 (m_{f_1(1285)}^2+m_{f_1(1400)}^2)(4m^2_{K_{1A}}-m_{a_1}^2)-3m_{f_1(1285)}^2m_{f_1(1400)}^2
=8m^4_{K_{1A}}-8m^2_{K_{1A}}m^2_{a_1}+3m_{a_1}^4.
 \en
This relation is satisfied for $^3P_1$ octet mesons, but only
approximately for $^1P_1$ states. Anyway, we shall use the mass
relation (\ref{eq:GMOkubo}) to fix the magnitude of the mixing
angle and (\ref{eq:GMOkubolinear}) to fix its sign.

Since $K^*\ov K$ and $K\ov K\pi$ are the dominant modes of
$f_1(1420)$ whereas $f_0(1285)$ decays mainly to the $4\pi$
states, this suggests that the quark content is primarily $s\bar
s$ for $f_1(1420)$ and $n\bar n$ for $f_1(1285)$. This may
indicate that $\theta_{^3P_1}=28^\circ$ is slightly preferred.
However, $\theta_{^3P_1}=53^\circ$ is equally acceptable.

Similarly, for $1^1P_1$ states, $h_1(1170)$ and $h_1(1380)$ may be
mixed in terms of the pure octet $h_8$ and singlet $h_1$,
 \begin{eqnarray}
 |h_1(1170)\rangle = |h_1\rangle\cos\theta_{^1P_1}+|h_8\rangle\sin\theta_{^1P_1},
 \quad
 |h_1(1380)\rangle = -|h_1\rangle\sin\theta_{^1P_1} +|h_8\rangle\cos\theta_{^1P_1} \,.
 \end{eqnarray}
Again from the Gell-Mann-Okubo mass formula, we obtain
\begin{equation}
\cos^2\theta_{^1P_1}
=\frac{4m_{K_{1B}}^2-m_{b_1}^2-3m_{h_1(1170)}^2}
{3\bigg(m_{h_1(1380)}^2-m_{h_1(1170)}^2\bigg)}\,,\label{eq:GMOkubo4}
\end{equation}
where
 \begin{equation}
 m_{K_{1B}}^2 =
 m_{K_1(1400)}^2 \sin^2\theta_{K_1} + m_{K_1(1270)}^2 \cos^2\theta_{K_1} \,.\label{eq:GMOkubo5}
 \end{equation}
We obtain $\theta_{^1P_1}^{\rm quad}=-18.1^\circ\,(25.2^\circ)$ and
$\theta_{^1P_1}^{\rm lin}=23.8^\circ\,(-18.3^\circ)$ for
$\theta_{K_1}=-37^\circ\,(-58^\circ)$, where the sign of the mixing
angle is determined from the mass relation
\begin{equation}
 \tan\theta_{^1P_1}
=\frac{4m_{K_{1B}}^2-m_{b_1}^2-3m_{h_1(1170)}^2}
{2\sqrt{2}(m_{b_1}^2-m_{K_{1B}}^2)}\,.
\end{equation}


\subsection{Decay constants}
 Decay constants of pseudoscalar and axial-vector mesons are defined as
 \be \label{eq:decayc}
 && \la P(p)|\bar q_2\gamma_\mu\gamma_5 q_1|0\ra=-if_P q_\mu, \quad
 \la ^{3(1)}P_1(p,\lambda)|\bar q_2\gamma_\mu\gamma_5 q_1|0\ra=if_{^3P_1(^1P_1)} m_{^3P_1(^1P_1)}\epsilon^{(\lambda)*}_\mu.
 \en
For axial-vector mesons, the transverse decay constant is
defined via the tensor current by
\begin{equation}
   \langle ^{3(1)}P_1(p,\lambda) |\bar q_2 \sigma^{\mu\nu}\gamma_5q_1 |0\rangle
 =-f_{^{3(1)}P_1}^\perp (\epsilon_{(\lambda)}^{*\mu} p^\nu -
\epsilon_{(\lambda)}^{*\nu} p^\mu)\,,
   \label{eq:tensor-1p1-2}
\end{equation}
or
 \be
 \langle ^{3(1)}P_1(p,\lambda) |\bar q_2 \sigma^{\mu\nu}q_1|0\rangle
 =-if_{^{3(1)}P_1}^\perp \,\epsilon_{\mu\nu\alpha\beta}
  \epsilon_{(\lambda)}^{*\alpha} p^\beta,
 \en
where we have applied the identity
$\sigma_{\alpha\beta}\gamma_5=-{i\over
2}\epsilon_{\alpha\beta\mu\nu}\sigma^{\mu\nu}$ with the sign
convention $\epsilon_{0123}=1$. Since the tensor current is not
conserved, the transverse decay constant $f^\perp$ is scale
dependent. Because of charge conjugation invariance, the decay
constant of the $^1P_1$ non-strange neutral meson $b_1^0(1235)$ must
be zero. In the isospin limit, the decay constant of the charged
$b_1$ vanishes due to the fact that the $b_1$ has even $G$-parity
and that the relevant weak axial-vector current is odd under $G$
transformation. Hence, $f_{b^\pm_1}$ is very small in reality. Note
that the matrix element of the pseudoscalar density vanishes,
$\langle ^{3(1)}P_1(p,\vp)|\bar q_2\gamma_5 q_1|0\ra=0$, which can
be seen by applying the equation of motion. As for the strange axial
vector mesons, the $^3P_1$ and $^1P_1$ states transfer under charge
conjunction as
 \be
 M_a^b(^3P_1) \to M_b^a(^3P_1), \qquad M_a^b(^1P_1) \to
 -M_b^a(^1P_1),~~~(a,b=1,2,3).
 \en
Since the weak axial-vector current transfers as $(A_\mu)_a^b\to
(A_\mu)_b^a$ under charge conjugation, it is clear that
$f_{^1P_1}=0$ in the SU(3) limit \cite{Suzuki}. By the same token,
the decay constant $f^\perp_{^3P_1}$ vanishes in the SU(3) limit.
Note for scalar mesons, their decay constants also vanish in the
same limit, which can be easily seen by applying equations of motion
to obtain
  \be \label{eq:Seom}
 m_S^2f_S=\,i(m_1-m_2)\la 0|\bar q_1q_2|S\ra,
 \en
with $m_i$ being the mass of the quark $q_i$.

The $a_1(1260)$ decay constant $f_{a_1}=238\pm10$ MeV obtained
using the QCD sum rule method \cite{YangNP} is similar to the
$\rho$ meson one, $f_\rho\approx$ 216 MeV. This means that the
$a_1(1260)$ can be regarded as the scalar partner of the $\rho$, as
it should be. To compute the decay constant $f_{b_1}$ for the
charged $b_1$, one needs to specify the $u$ and $d$ quark mass
difference in the model calculation. In the covariant light-front
quark model \cite{CCH}, if we increase the constituent $d$ quark
mass by an amount of $5\pm2$ MeV relative to the $u$ quark one, we
find $f_{b_1}=0.6\pm0.2$ MeV which is highly suppressed. As we
shall see below, the decay constant $f_{b_1}$ is related to the
transverse one $f_{b_1}^\perp$ by the relation [see Eq.
(\ref{eq:f1p1})]
 \be
 f_{b_1}=f_{b_1}^\perp(\mu) a_0^{\parallel,b_1}(\mu),
 \en
where $a_0^{\parallel,b_1}$ is the zeroth Gegenbauer moment of
$\Phi_\parallel^{b_1}$ to be defined later. The quantities
$f_{b_1}^\perp$ and $a_0^{\parallel,b_1}$ can be calculated in the
QCD sum rule approach with the results $f_{b_1}^\perp=(180\pm8)$
MeV \cite{YangNP} (cf. Table \ref{tab:decayc}) and
$a_0^{\parallel,b_1}=0.0028\pm0.0026$ for $b_1^-$  at $\mu=1$ GeV.
(Note that for $b_1^+$, $a_0^{\parallel,b_1}$ has an opposite sign
due to $G$-parity.) Again, $f_{b_1}$ is very small, of order 0.5
MeV, in agreement with the estimation based on the light-front
quark model. In \cite{Nardulli07}, the decay constants of $a_1$ and
$b_1$ are derived using the $K_{1A}-K_{1B}$ mixing angle
$\theta_{K_1}$ and SU(3) symmetry: $(f_{b_1},f_{a_1})=(74,~215)$
MeV for $\theta_{K_1}=32^\circ$ and $(-28,223)$ MeV for
$\theta_{K_1}=58^\circ$. It seems to us that the $b_1$ decay
constant derived in this manner is too big.

Introducing the decay constants $f_{f_1(1285)}^q$ and
$f_{f_1(1420)}^q$ by
\begin{eqnarray}
\langle 0| \bar q\gamma_\mu \gamma_5 q |
f_1(1285)(P,\lambda)\rangle
 &=& -i m_{f_1(1285)} f_{f_1(1285)}^q \epsilon_\mu^{(\lambda)}\,,\label{eq:decay-def1}\\
\langle 0| \bar q\gamma_\mu \gamma_5 q |
f_1(1420)(P,\lambda)\rangle
 &=& -i m_{f_1(1420)} f_{f_1(1420)}^q \epsilon_\mu^{(\lambda)}\,, \label{eq:decay-def2}
\end{eqnarray}
we obtain
\begin{eqnarray}
f_{f_1(1285)}^u &=&
\frac{f_{f_1}}{\sqrt{3}}\frac{m_{f_1}}{m_{f_1(1285)}}\cos\theta_{^3P_1}
 +  \frac{f_{f_8}}{\sqrt{6}}\frac{m_{f_8}}{m_{f_1(1285)}}\sin\theta_{^3P_1}
 = 172\pm 23 ~(178\pm 22)~{\rm MeV} \,,~~~~~ \label{eq:decay-f1-1}\\
f_{f_1(1285)}^s &=&
\frac{f_{f_1}}{\sqrt{3}}\frac{m_{f_1}}{m_{f_1(1285)}}\cos\theta_{^3P_1}
 -  \frac{2 f_{f_8}}{\sqrt{6}}\frac{m_{f_8}}{m_{f_1(1285)}}\sin\theta_{^3P_1}
 = -72\pm 13~(29\pm 18)~{\rm MeV} \,,  \label{eq:decay-f1-2}\\
f_{f_1(1420)}^u &=&
-\frac{f_{f_1}}{\sqrt{3}}\frac{m_{f_1}}{m_{f_1(1420)}}\sin\theta_{^3P_1}
 +  \frac{f_{f_8}}{\sqrt{6}}\frac{m_{f_8}}{m_{f_1(1420)}}\cos\theta_{^3P_1}
 = -55 \pm 10~(23\pm 11)~{\rm MeV} \,,  \label{eq:decay-f1-3}\\
f_{f_1(1420)}^s &=&
-\frac{f_{f_1}}{\sqrt{3}}\frac{m_{f_1}}{m_{f_1(1420)}}\sin\theta_{^3P_1}
 -  \frac{2 f_{f_8}}{\sqrt{6}}\frac{m_{f_8}}{m_{f_1(1420)}}\cos\theta_{^3P_1}  \nonumber\\
 &=& -219\pm 27~(-230\pm 26)~{\rm MeV}\,, \label{eq:decay-f1-4}
\end{eqnarray}
corresponding to $\theta_{^3P_1}=53.2^\circ (27.9^\circ)$, where
we have used the QCD sum rule results for $f_{f_1}$ and $f_{f_8}$
\cite{YangNP} (see Table \ref{tab:decayc}).

The decay constants for $K_1(1270)$ and $K_1(1400)$ defined by
(with $\bar q=\bar u$ or $\bar d$)
 \begin{eqnarray}\label{eq:k1-1}
 \langle 0 |\bar q\gamma_\mu \gamma_5 s |K_{1}(1270)(P,\lambda)\rangle
 &=& -i \, f_{K_{1}(1270)}\,
 m_{K_{1}(1270)}\,\epsilon_\mu^{(\lambda)},
 \end{eqnarray}
and
 \begin{eqnarray}\label{eq:k1-2}
\langle 0 |\bar q\gamma_\mu \gamma_5
s|K_{1}(1400)(P,\lambda)\rangle
  &=& -i \, f_{K_{1}(1400)}\,
  m_{K_{1}(1400)}\,\epsilon_\mu^{(\lambda)}
 \end{eqnarray}
are related to $f_{K_{1A}}$ and $f_{K_{1B}}$ by
 \be
 f_{K_{1}(1270)} &=& {1\over m_{K_{1}(1270)}}(f_{K_{1A}} m_{K_{1A}} \sin{\theta_{K_1}}
    + f_{K_{1B}} m_{K_{1B}} \cos{\theta_{K_1}}), \non \\
  f_{K_{1}(1400)} &=& {1\over m_{K_{1}(1400)}}(f_{K_{1A}} m_{K_{1A}} \cos{\theta_{K_1}}
    - f_{K_{1B}} m_{K_{1B}} \sin{\theta_{K_1}}).
 \en
Just as the previous $b_1$ case, the decay constant $f_{K_{1B}}$ is
related to the transverse one $f_{K_{1B}}^\perp$ by the relation
 $f_{K_{1B}}=f_{K_{1B}}^\perp(\mu) a_0^{\parallel,K_{1B}}(\mu)$.
If we apply the QCD sum rule results for $f_{K_{1A}}$,
$f_{K_{1B}}^\perp$ (see Table \ref{tab:decayc}) and
$a_0^{\parallel,K_{1B}}$ (cf. Table \ref{tab:Gegenbauer}), we will
obtain
 \be
&&  f_{K_1(1270)}=-137\pm15~{\rm MeV}, \qquad
f_{K_1(1400)}=199\pm10~{\rm MeV}, \quad~{\rm for\
}\theta_{K_1}=-37^\circ, \non \\
&& f_{K_1(1270)}=-207\pm7~{\rm MeV}, \qquad
f_{K_1(1400)}=141\pm14~{\rm MeV}, \qquad {\rm for\
}\theta_{K_1}=-58^\circ.
 \en
However, we would like to make two remarks. First, we do have the
experimental information on the decay constant of
$K_1(1270)$.\footnote{The large experimental error with the
$K_1(1400)$ production in the $\tau$ decay, namely, $\B(\tau^-\to
K_1^-(1400)\nu_\tau) = (1.7\pm2.6)\times 10^{-3}$ \cite{PDG}, does
not provide sensible information for the $K_1(1400)$ decay
constant.}
From the measured branching ratio of $\tau\to K_1^-(1270)\nu_\tau$
by ALEPH \cite{ALEPH}, $\B(\tau^-\to K_1^-(1270)\nu_\tau) =
(4.7\pm1.1)\times 10^{-3}$, the decay constant of $K_1(1270)$ is
extracted to be \cite{ChengDAP}
 \be
 \left|f_{K_1(1270)}\right|=175\pm 19~{\rm MeV},
 \en
where use has been made of the formula
 \be
 \Gamma(\tau\to K_1\nu_\tau)={G_F^2\over
 16\pi}|V_{us}|^2\,f_{K_1}^2{(m_\tau^2+2m_{K_1}^2)(m_\tau^2-m_{K_1}^2)^2\over
 m_\tau^3}.
 \en
Second, as pointed out in \cite{CCH}, the decay constants of
$^3P_1$ have opposite signs to that of $^1P_1$ in the covariant
light-front quark model. The large error with the QCD sum rule
result of $a_0^{\parallel,K_{1B}}=0.14\pm0.15$ is already an
indication of possible large sum rule uncertainties in this
quantity.

In order to reduce the theoretical uncertainties with the $K_1$
decay constant, we shall use the experimental value of
$f_{K_1(1270)}$ to fix the input parameters $\beta_{K_{1A}}$ and
$\beta_{K_{1B}}$ appearing in the Gaussian-type wave function in
the covariant quark model \cite{CCH}. We obtain
 \be
 \beta_{K_{1A}}=\beta_{K_{1B}}=\cases{0.375\,{\rm GeV}; &
 for~$\theta_{K_1}=-37^\circ$, \cr 0.313\,{\rm GeV}; &
 for~$\theta_{K_1}=-58^\circ$,  }
 \en
and
 \be
&&  f_{K_{1A}}=293~{\rm MeV}, \qquad f_{K_{1B}}=15~{\rm MeV}, \qquad
{\rm for\ }\theta_{K_1}=-37^\circ, \non \\ && f_{K_{1A}}=207~{\rm
MeV}, \qquad f_{K_{1B}}=12~{\rm MeV}, \qquad {\rm for\
}\theta_{K_1}=-58^\circ.
 \en
Therefore, we have
\begin{eqnarray} \label{eq:fK1}
f_{K_{1}(1270)} &=& -175  \pm 11~ {\rm MeV} \,, \quad
f_{K_{1}(1400)} = 235 \pm 12~ {\rm MeV} \,,
 \qquad {\rm for\ }\theta_{K_1}=-37^\circ, \non \\
f_{K_{1}(1270)} &=& -175 \pm 15~ {\rm MeV}, \quad f_{K_{1}(1400)} =
112 \pm 12~ {\rm MeV} \,,
 \qquad~ {\rm for\ }\theta_{K_1}=-58^\circ.
\end{eqnarray}

\begin{table}[t]
\caption{ Summary of the decay constants $f_{^3P_1}$ and
$f^\perp_{^1P_1}(1\,{\rm GeV})$ in units of MeV obtained from QCD
sum rule methods \cite{YangNP}.} \label{tab:decayc}
 \vspace{0.1cm}
\begin{center}
{\tabcolsep=0.58cm\begin{tabular}{|c|c|c|c|c|} \hline
$^3P_1$ & $a_1(1260)$ & $f_1$ & $f_8$ & $K_{1A}$\\
\hline
$f_{^3P_1}$ & $238\pm10$ & $245\pm13$ & $239\pm13$ & $250\pm13$\\
\hline
 \end{tabular}}
{\tabcolsep=0.58cm\begin{tabular}{|c|c|c|c|c|} \hline
$^1P_1$ & $b_1(1235)$ & $h_1$ & $h_8$ & $K_{1B}$\\
\hline
$f_{^1P_1}^\perp$ & $180\pm8$ & $180\pm12$ & $190\pm10$ & $190\pm10$\\
\hline
\end{tabular}}
\end{center}
\end{table}

In complete analogy to the discussion for $1^3P_1$ states, we
introduce the tensor couplings for $1^1P_1$ states
  \begin{eqnarray}
\langle 0| \bar q\sigma_{\mu\nu} q | h_1(1170)(P,\lambda)\rangle
 &=& i f_{h_1(1170)}^{\perp,q}\,\epsilon_{\mu\nu\alpha\beta}
  \epsilon_{(\lambda)}^\alpha P^\beta\,, \label{eq:decay-def3}\\
\langle 0| \bar q\sigma_{\mu\nu} q | h_1(1380)(P,\lambda)\rangle
 &=& i f_{h_1(1380)}^{\perp,q}\,\epsilon_{\mu\nu\alpha\beta}
  \epsilon_{(\lambda)}^\alpha P^\beta\,, \label{eq:decay-def4}
\end{eqnarray}
and then obtain
\begin{eqnarray}
f_{h_1(1170)}^{\perp,u} &=&
\frac{f_{h_1}^\perp}{\sqrt{3}}\cos\theta_{^1P_1}
 +  \frac{f_{h_8}^\perp}{\sqrt{6}}\sin\theta_{^1P_1}
 = 75\pm 8 ~(127\pm 7)~{\rm MeV} \,,~~~~~  \label{eq:decay-h1-1}\\
f_{h_1(1170)}^{\perp,s} &=&
\frac{f_{h_1}^\perp}{\sqrt{3}}\cos\theta_{^1P_1}
 -  \frac{2 f_{h_8}^\perp}{\sqrt{6}}\sin\theta_{^1P_1}
 = 147\pm 8~(28\pm 10)~{\rm MeV} \,,  \label{eq:decay-h1-2}\\
f_{h_1(1380)}^{\perp,u} &=& -\frac{f_{h_1}^\perp}{\sqrt{3}}
\sin\theta_{^1P_1}
 +  \frac{f_{h_8}^\perp}{\sqrt{6}} \cos\theta_{^1P_1}
 = 106\pm 5~(26\pm 7)~{\rm MeV} \,,  \label{eq:decay-h1-3}\\
f_{h_1(1380)}^{\perp,s} &=& -\frac{f_{h_1}^\perp}{\sqrt{3}}
\sin\theta_{^1P_1}
 -  \frac{2 f_{h_8}^\perp}{\sqrt{6}} \cos\theta_{^1P_1}
 = -115\pm 9~(-185\pm 10)~{\rm MeV}\,,  \label{eq:decay-h1-4}
\end{eqnarray}
corresponding to $\theta_{^1P_1}=-18.1^\circ (25.2^\circ)$ where
we have used the QCD sum rule results for $f_{h_1}^\perp$ and
$f_{h_8}^\perp$ given in Table \ref{tab:decayc} \cite{YangNP}.

As for strange axial-vector mesons, we have (with $\bar q\equiv
\bar u, \bar d$)
 \begin{eqnarray}\label{eq:K1-1270}
 \langle 0 |\bar q\sigma_{\mu\nu}s |K_{1}(1270)(p,\lambda)\rangle
 &=& i f_{K_{1}(1270)}^\perp
 \,\epsilon_{\mu\nu\alpha\beta} \epsilon_{(\lambda)}^\alpha
 p^\beta\nonumber\\
 &=& i (f_{K_{1A}}^\perp \sin {\theta_{K_1}}
    + f_{K_{1B}}^\perp\cos {\theta_{K_1}})
 \,\epsilon_{\mu\nu\alpha\beta} \epsilon_{(\lambda)}^\alpha p^\beta
 \end{eqnarray}
and
 \begin{eqnarray}\label{eq:K1-1400}
 \langle 0 |\bar q\sigma_{\mu\nu}s |K_{1}(1400)(p,\lambda)\rangle
 &=& i f_{K_{1}(1400)}^\perp
 \,\epsilon_{\mu\nu\alpha\beta} \epsilon_{(\lambda)}^\alpha
 p^\beta\nonumber\\
 &=& i (f_{K_{1A}}^\perp \cos {\theta_{K_1}}
     - f_{K_{1B}}^\perp \sin {\theta_{K_1}})
 \,\epsilon_{\mu\nu\alpha\beta}\epsilon_{(\lambda)}^\alpha
 p^\beta.
 \end{eqnarray}
As will be shown in Sec. II.D below, the decay constants
$f_{K_{1A}}$ and $f_{K_{1A}}^\perp$ are related via
 \be
 f_{K_{1A}}^\perp(\mu)=f_{K_{1A}}a_0^{\perp,K_{1A}}(\mu).
 \en
From Tables \ref{tab:decayc} and \ref{tab:Gegenbauer}, we obtain
(at the scale $\mu=1$~GeV)
\begin{eqnarray}
f_{K_{1}(1270)}^\perp &=& 140  \pm 22~ {\rm MeV} \,,\quad
f_{K_{1}(1400)}^\perp = 130 \pm 25~ {\rm MeV} \,,
 \qquad {\rm for \ }\theta_{K_1}=-37^\circ,  \non \\
f_{K_{1}(1270)}^\perp &=& ~ 84  \pm 25~ {\rm MeV} \,, \quad
f_{K_{1}(1400)}^\perp = 172 \pm 21~ {\rm MeV} \,,
 \qquad {\rm for\ }\theta_{K_1}=-58^\circ.
\end{eqnarray}

\subsection{Form factors}
The form factors for the $\ov B\to A$ and $\ov B\to P$ transitions
are defined as
\begin{eqnarray} \label{eq:FF}
\langle{A}(p, \lambda)|A_\mu|{\overline B} (p_B)\rangle
 &=& i \frac{2}{m_B - m_{A}} \epsilon_{\mu\nu\alpha\beta} \epsilon_{(\lambda)}^{*\nu}
 p_B^\alpha p^{\beta} A^{BA}(q^2),
\nonumber \\
 \langle A (p,\lambda)|V_\mu|{\overline B}(p_B)\rangle
 &=& - \Bigg\{ (m_B - m_{A}) \epsilon^{(\lambda)*}_{\mu} V_1^{BA}(q^2)
 - (\epsilon^{(\lambda)*} \cdot p_B)
(p_B + p)_\mu \frac{V_2^{BA}(q^2)}{m_B - m_{A}}
\nonumber \\
&& - 2 m_{A} \frac{\epsilon^*_{(\lambda)} p_B}{q^2} q^\mu
\left[V_3^{BA}(q^2) - V_0^{BA}(q^2)\right]\Bigg\}, \nonumber\\
 \langle P(p)|V_\mu|\overline B(p_B)\rangle &=&
 \left[ (p_B+p)_\mu-{m_B^2-m_P^2\over q^2}\,q_ \mu\right]
 F_1^{BP}(q^2)+{m_B^2-m_P^2\over q^2}q_\mu\,F_0^{BP}(q^2), \nonumber \\
 \end{eqnarray}
where $q = p_B - p$, $V_3^{BA}(0) = V_0^{BA}(0)$,
$F_1^{BP}(0)=F_0^{BP}(0)$ and
 \begin{eqnarray}
 V_3^{BA}(q^2)= \frac{m_B - m_{A}}{2 m_{A}} V_1^{BA}(q^2) - \frac{m_B
 + m_{A}}{2 m_{A}} V_2^{BA}(q^2).
 \end{eqnarray}

In the literature the decay constant and the form factors of the
axial vector mesons are often defined in different manner. For
example, in \cite{CCH} they are defined as\footnote{Since the convention
$\epsilon^{1234}=+1$ is adopted in \cite{CCH} while $\epsilon_{1234}=+1$
is used in the present work, we have put an additional minus
sign for the matrix element $\langle{A}(p, \lambda)|A_\mu|{\overline B}
(p_B)\rangle$.}
 \be \label{eq:deccCCH}
 \la A(p,\lambda)|A_\mu|0\ra &=& f_{A}m_{A}\epsilon_{\mu}^{(\lambda)*},
 \en
 \be \label{eq:FFCCH}
 \langle{A}(p, \lambda)|A_\mu|{\overline B} (p_B)\rangle
 &=& -\frac{2}{m_B - m_{A}} \epsilon_{\mu\nu\alpha\beta} \epsilon_{(\lambda)}^{*\nu}
 p_B^\alpha p^{\beta} A^{BA}(q^2),
\nonumber \\
  \langle A (p,\lambda)|V_\mu|{\overline B}(p_B)\rangle
 &=& -i \Bigg\{ (m_B - m_{A}) \epsilon^{(\lambda)*}_{\mu} V_1^{BA}(q^2)
 - (\epsilon^{*}_{(\lambda)} p_B)
(p_B + p)_\mu \frac{V_2^{BA}(q^2)}{m_B - m_{A}} \non \\
&&  -2 m_{A} \frac{\epsilon^{*}_{(\lambda)} p_B}{q^2} q^\mu
\left[V_3^{BA}(q^2) - V_0^{BA}(q^2)\right]\Bigg\}.
 \en
It has been checked in the covariant light-front quark model that
the form factors $V_{0,1,2}^{B\,^3P_1}(q^2)$ and
$A^{B\,^3P_1}(q^2)$ defined in Eq. (\ref{eq:FFCCH}) are indeed
positively defined. We would like to ask if the $\ov B\to\, ^3P_1$
transition form factors defined in Eq. (\ref{eq:FF}) are also
positively defined. This can be checked by considering the
factorizable amplitudes for the decay $\ov B\to AP$
 \begin{eqnarray} \label{eq:X}
  X^{(\bar B A, P)}
 &=& \langle P(q)| (V-A)_\mu|0\rangle \langle A(p)| (V-A)^\mu|\overline
 B(p_B)\rangle, \non \\
  X^{(\bar B P, A)}
 &=& \langle A(q)| (V-A)_\mu|0\rangle \langle P(p)| (V-A)^\mu|\overline
 B(p_B)\rangle.
 \end{eqnarray}
We obtain
 \be \label{eq:XAP}
 X^{(\bar B A, P)}
 = - 2 i f_P m_{A} V_0^{BA}(q^2) (\epsilon^{*}_{(\lambda)} p_B)\,,  \qquad
  X^{(\bar B P, A)}=
 - 2 i f_A m_{A} F_1^{BP}(q^2) (\epsilon^{*}_{(\lambda)} p_B)\,,
 \end{eqnarray}
from Eqs. (\ref{eq:decayc}) and (\ref{eq:FF}) and
 \be
 X^{(\bar B A, P)}
 =  2  f_P m_{A} V_0^{BA}(q^2) (\epsilon^{*}_{(\lambda)} p_B)\,,  \qquad
  X^{(\bar B P, A)}=
 - 2  f_A m_{A} F_1^{BP}(q^2) (\epsilon^{*}_{(\lambda)} p_B)\,,
 \end{eqnarray}
from Eqs. (\ref{eq:deccCCH}) and (\ref{eq:FFCCH}). Since $f_A$ for
the $^3P_1$ meson is negative in the light-front model calculation
(see Eq. (2.23) and Table III in \cite{CCH}), the relative sign
between $X^{(\bar B A, P)}$ and $X^{(\bar B P, A)}$ is positive.
This means that the relative sign in Eq. (\ref{eq:XAP}) is also
positive provided that the decay constant $f_A$ and the form
factor $V_0^{BA}$ defined in Eqs. (\ref{eq:decayc}) and
(\ref{eq:FF}), respectively, are of the same sign. Indeed, it is
found in \cite{Yanga1} that if $f_A$ is chosen to be positive for
the $a_1(1260)$ meson, the form factor $V_0^{Ba_1}$ is indeed
positive according to the sum rule calculation.


\begin{table}[t]
\caption{Form factors for
$B\to\pi,K,a_1(1260),b_1(1235),K_{1A},K_{1B}$ transitions obtained
in the covariant light-front model \cite{CCH} are fitted to the
3-parameter form Eq. (\ref{eq:FFpara}) except for the form factor
$V_2$ denoted by $^{*}$ for which the fit formula Eq.
(\ref{eq:FFpara1}) is used.} \label{tab:LFBtopi}
\begin{ruledtabular}
\begin{tabular}{| c c c c c || c c c c c |}
~~~$F$~~~~~
    & $F(0)$~~~~~
    & $F(q^2_{\rm max})$~~~~
    &$a$~~~~~
    & $b$~~~~~~
& ~~~ $F$~~~~~
    & $F(0)$~~~~~
    & $F(q^2_{\rm max})$~~~~~
    & $a$~~~~~
    & $b$~~~~~~
 \\
    \hline
$F^{B\pi}_1$
    & $0.25$
    & $1.16$
    & 1.73
    & 0.95
& $F^{B\pi}_0$
    & 0.25
    & 0.86
    & 0.84
    & $0.10$
    \\
$F^{BK}_1$
    & $0.35$
    & $2.17$
    & 1.58
    & 0.68
& $F^{BK}_0$
    & 0.35
    & 0.80
    & 0.71
    & $0.04$
    \\
$A^{Ba_1}$
    & $0.25$
    & $0.76$
    & 1.51
    & 0.64
&$V^{Ba_1}_0$
    & $0.13$
    & $0.32$
    &  1.71
    &  1.23
    \\
$V^{Ba_1}_1$
    & $0.37$
    & $0.42$
    & $0.29$
    & 0.14
&$V^{Ba_1}_2$
    & $0.18$
    & $0.36$
    & $1.14$
    & $0.49$
    \\
$A^{Bb_1}$
    & $-0.10$
    & $-0.23$
    & 1.92
    & 1.62
&$V^{Bb_1}_0$
    & $-0.39$
    & $-0.98$
    & 1.41
    & 0.66
    \\
$V^{Bb_1}_1$
    & $0.18$
    & $0.36$
    & $1.03$
    & 0.32
&$V^{Bb_1}_2$
    & $0.03^*$
    & $-0.15^*$
    & $2.13^*$
    & $2.39^*$
    \\
$A^{BK_{1A}}$
    & $0.26$
    & $0.69$
    & 1.47
    & 0.59
&$V^{BK_{1A}}_0$
    & $0.14$
    & $0.31$
    &  1.62
    &  1.14
    \\
$V^{BK_{1A}}_1$
    & $0.39$
    & $0.42$
    & $0.21$
    & 0.16
&$V^{BK_{1A}}_2$
    & $0.17$
    & $0.30$
    & $1.02$
    & $0.45$
    \\
$A^{BK_{1B}}$
    & $-0.11$
    & $-0.25$
    & 1.88
    & 1.53
&$V^{BK_{1B}}_0$
    & $-0.41$
    & $-0.99$
    & $1.40$
    & $0.64$
    \\
$V^{BK_{1B}}_1$
    & $-0.19$
    & $-0.35$
    & 0.96
    & 0.30
&$V^{BK_{1B}}_2$
    & $0.05^*$
    & $0.16^*$
    & $1.78^*$
    & $2.12^*$
    \\
\end{tabular}
\end{ruledtabular}
\end{table}

\begin{table}[t]
\caption{Form factors for $B\to
a_1(1260),b_1(1235),K_{1A},K_{1B},f_1,f_8,h_1,h_8$ transitions at
$q^2=0$ obtained in the framework of the light-cone sum rule
approach \cite{YangFF}. Uncertainties arise from the Borel window
and the input parameters.} \label{tab:SRFF}
\begin{tabular}{| l r |  l r |}
\hline \hline
    \qquad $V^{Ba_1}_0(0)$\qquad
    & \qquad $0.303^{+0.022}_{-0.035}$ \qquad
    & \qquad $V^{Bb_1}_0(0)$ \qquad
    & \qquad$-0.356^{+0.039}_{-0.033}$ \qquad
    \\
    \qquad $V^{BK_{1A}}_0(0)$\qquad
    & \qquad $0.316^{+0.048}_{-0.042}$ \qquad
    & \qquad $V^{BK_{1B}}_0(0)$ \qquad
    & \qquad$-0.360^{+0.030}_{-0.028}$ \qquad
    \\
    \qquad $V^{Bf_1}_0(0)$\qquad
    & \qquad $0.181^{+0.018}_{-0.021}$ \qquad
    & \qquad $V^{Bh_1}_0(0)$\qquad
    & \qquad $-0.214^{+0.021}_{-0.012}$ \qquad
    \\
    \qquad $V^{Bf_8}_0(0)$ \qquad
    & \qquad$0.124^{+0.015}_{-0.004}$ \qquad
    & \qquad $V^{Bh_8}_0(0)$ \qquad
    & \qquad$ -0.158^{+0.016}_{-0.018}$ \qquad
    \\ \hline \hline
\end{tabular}
\end{table}

The form factors for $B\to\pi,K,a_1(1260),b_1(1235),K_{1A},K_{1B}$
transitions have been calculated in  the relativistic covariant
light-front (CLF) quark model  \cite{CCH} (Table
\ref{tab:LFBtopi})\footnote{As explained in the footnote before Eq.
(\ref{eq:RK1gamma}), we need to put additional minus signs to the
$B\to\, ^1P_1$ form factors in Table \ref{tab:LFBtopi} since in the
convention of the present work, the form factors $V_i^{B\to ^1P_1}$
and $V_i^{B\to ^3P_1}$ have different signs.}
and in the framework of the light-cone sum rule (LCSR) approach
\cite{YangFF}. In the CLF model, the momentum dependence of the
physical form factors is determined by first fitting the form
factors obtained in the spacelike region to a 3-parameter function
in $q^2$ and then analytically continuing them to the timelike
region. Some of the $V_2(q^2)$ form factors in $P\to A$
transitions are fitted to a different 3-parameter form so that the
fit parameters are stable within the chosen $q^2$ range.

Except for the form factor $V_2$ to be discussed below, it is
found in \cite{CCH} that the momentum dependence of form factors
in the spacelike region can be well parameterized and reproduced
in the three-parameter form:
 \be \label{eq:FFpara}
 F(q^2)=\,{F(0)\over 1-a(q^2/m_{B}^2)+b(q^2/m_{B}^2)^2}\,,
 \en
for $B\to M$ transitions. The parameters $a$, $b$ and $F(0)$ are
first determined in the spacelike region. We then employ this
parametrization to determine the physical form factors at $q^2\geq
0$. In practice, these parameters
are generally insensitive to the $q^2$ range to be fitted except
for the form factor $V_2(q^2)$ in $B\to\, ^1P_1$ transitions. The
corresponding parameters $a$ and $b$ are rather sensitive to the
chosen range for $q^2$. This sensitivity is attributed to the fact
that the form factor $V_2(q^2)$ approaches to zero at very large
$-|q^2|$ where the three-parameter parametrization
(\ref{eq:FFpara}) becomes questionable. To overcome this
difficulty, we will fit this form factor to the form
 \be \label{eq:FFpara1}
 F(q^2)=\,{F(0)\over (1-q^2/m_{B}^2)[1-a(q^2/m_{B}^2)+b(q^2/m_{B}^2)^2]}
 \en
and achieve a substantial improvement \cite{CCH}.

Momentum dependence of the form factors calculated using the LCSR
method is not shown in Table \ref{tab:SRFF}.
Since the pseudoscalar mesons considered in the present work are
the light pion and the kaon, the form-factor $q^2$ dependence can be
neglected for our purposes.

\begin{table}[h]
\caption{$B\to a_1(1260)$ transition form factor $V_0^{Ba_1}$ at
$q^2=0$ in various models, where the QSR1 is the traditional QCD
sum rule approach and the QSR2 is the light-cone sum rule
approach.} \label{tab:FFa1}
\begin{ruledtabular}
\begin{tabular}{c c c c c c}
 & CLF \cite{CCH} & ISGW2 \cite{ISGW2} & CQM \cite{Deandrea} & QSR1 \cite{Aliev} & QSR2 \cite{YangFF} \\
 \hline
 $V_0^{Ba_1}(0)$ & 0.13 & 1.01 & 1.20 & $0.23\pm0.05$ &
 $0.303^{+0.022}_{-0.035}$ \\
\end{tabular}
\end{ruledtabular}
\end{table}

In principle, the experimental measurements of $\ov B^0\to
a_1^+\pi^-$ and $\ov B^0\to b_1^+\pi^-$ will enable us to test the
form factors $V_0^{Ba_1}$ and $V_0^{Bb_1}$ respectively. There are
several existing model calculations for $B\to a_1$ form factors: one
in a quark-meson model (CQM) \cite{Deandrea}, one in the ISGW2 model
\cite{ISGW2}, one in the light-front quark model \cite{CCH} and two
based on the QCD sum rule (QSR) \cite{Aliev,YangFF}. Predictions in
various models are summarized in Table \ref{tab:FFa1} and in general
they are quite different. For example, $V_0^{Ba_1}(0)$ obtained in
the quark-meson model, 1.20\,, is larger than the value of the
sum-rule prediction in \cite{Aliev}, $0.23\pm 0.05$.  If $a_1(1260)$
behaves as the scalar partner of the $\rho$ meson, $V_0^{Ba_1}$ is
expected to be similar to $A_0^{B\rho}$, which is of order 0.28 at
$q^2=0$ \cite{CCH}. Indeed, the sum rule calculation by one of us
(K.C.Y.) yields $V_0^{Ba_1}=0.303^{+0.022}_{-0.035}$\,. Therefore,
it appears to us that a magnitude of order unity for $V_0^{Ba_1}(0)$
as predicted by the ISGW2 model and CQM is very unlikely. The BaBar
measurement of $\ov B^0\to a_1^+\pi^-$ \cite{BaBara1pitime} favors a
value of $V_0^{Ba_1}(0)\approx 0.30$, which is very close the LCSR
result shown in Table \ref{tab:SRFF}.

Various $B\to A$ form factors also have been calculated in the
Isgur-Scora-Grinstein-Wise  (ISGW) model \cite{ISGW,ISGW2} based
on the nonrelativistic constituent quark picture. As pointed out
in \cite{CCH}, in general, the form factors at small $q^2$ in CLF
and ISGW models agree within 40\%. However, $F_0^{BD_0^*}(q^2)$
and $V_1^{BD_1^{1/2}}(q^2)$ have a very different $q^2$ behavior
in these two models as $q^2$ increases. Relativistic effects are
mild in $B\to D^{**}$ transitions but can manifest in
heavy-to-light transitions at maximum recoil. For example,
$V_0^{Ba_1}(0)$ is found to be 0.13 in the CLF model, while it is
as big as 1.01 in the ISGW2 model.

\subsection{Light-cone distribution amplitudes}
For an axial-vector meson, the chiral-even LCDAs are given by
\begin{eqnarray}
  &&\langle A(P,\lambda)|\bar q_1(y) \gamma_\mu \gamma_5 q_2(x)|0\rangle
  = i m_{A} \, \int_0^1
      du \,  e^{i (u \, p y + \bar u p x)}
   \left\{p_\mu \,
    \frac{\epsilon^{(\lambda)*} z}{p z} \, \Phi_\parallel(u)
         +\epsilon_{\perp\, \mu}^{(\lambda)*} \, g_\perp^{(a)}(u)
         \right\}, \nonumber\\ \label{eq:evendef1} \\
  &&\langle A(P,\lambda)|\bar q_1(y) \gamma_\mu
  q_2(x)|0\rangle
  = - i m_A \,\epsilon_{\mu\nu\rho\sigma} \,
      \epsilon^{*\nu}_{(\lambda)} p^{\rho} z^\sigma \,
    \int_0^1 du \,  e^{i (u \, p y +
    \bar u p x)} \,
       \frac{g_\perp^{(v)}(u)}{4}, \label{eq:evendef2}
\end{eqnarray}
with $u \, (\bar u=1-u)$ being the momentum fraction carried by
$q_1 (\bar q_2)$, and the chiral-odd LCDAs read
\begin{eqnarray}
  &&\langle A(P,\lambda)|\bar q_1(y) \sigma_{\mu\nu}\gamma_5 q_2(x)
            |0\rangle
  =  \int_0^1 du \, e^{i (u \, p y +
    \bar u p x)} \,
\Bigg\{(\epsilon^{(\lambda)*}_{\perp\mu} p_{\nu} -
  \epsilon_{\perp\nu}^{(\lambda)*}  p_{\mu}) \,
  \Phi_\perp(u)  \nonumber\\
&& \hspace*{+5cm}
  + \,\frac{m_A^2\,\epsilon^{(\lambda)*} z}{(p z)^2} \,
   (p_\mu z_\nu -
    p_\nu  z_\mu) \, h_\parallel^{(t)}(u)
\Bigg\},\label{eq:odddef1}\\
&&\langle A(P,\lambda)|\bar q_1(y) \gamma_5 q_2(x)
            |0\rangle
  =   m_{A}^2 \epsilon^{(\lambda)*} z\,\int_0^1 du \, e^{i (u \, p y +
    \bar u p x)}  \, \frac{h_\parallel^{(p)}(u)}{2}\,.\label{eq:odddef2}
\end{eqnarray}
Here, throughout the present discussion, we define $z\equiv y-x$
with $z^2=0$ and introduce the light-like vector
$p_\mu=P_\mu-m_A^2 z_\mu/(2 P z)$ with the meson's momentum
${P}^2=m_A^2$. Moreover, the meson polarization vector
$\epsilon_\mu$ has been decomposed into longitudinal
($\epsilon^{(\lambda)*}_{\parallel\mu}$) and transverse
($\epsilon^{(\lambda)*}_{\perp\mu}$) {\it projections} defined as
\begin{eqnarray}\label{eq:polprojectiors}
 && \epsilon^{(\lambda)*}_{\parallel\mu} \equiv
     \frac{\epsilon_{(\lambda)}^* z}{P z} \left(
      P_\mu-\frac{m_A^2}{P z} \,z_\mu\right), \qquad
 \epsilon^{(\lambda)*}_{\perp\mu}
        = \epsilon^{(\lambda)*}_\mu -\epsilon^{(\lambda)*}_{\parallel\mu}\,,
\end{eqnarray}
respectively. The LCDAs $\Phi_\parallel, \Phi_\perp$ are of
twist-2, and $g_\perp^{(v)}, g_\perp^{(a)}, h_\perp^{(t)},
h_\parallel^{(p)}$ of twist-3. Due to $G$-parity, $\Phi_\parallel,
g_\perp^{(v)}$ and $g_\perp^{(a)}$ are symmetric (antisymmetric)
with the replacement of $u\to 1-u$ for $^3P_1$ ($^1P_1$) states ,
whereas $\Phi_\perp, h_\parallel^{(t)}$ and $h_\parallel^{(p)}$
are antisymmetric (symmetric) in the SU(3) limit \cite{YangNP}. We
restrict ourselves to two-parton LCDAs with twist-3 accuracy.

Assuming that the axial-vector meson moves along the $z-$axis, the
derivation for the light-cone projection operator of an
axial-vector meson in the momentum space is in complete analogy to
the case of the vector meson. We separate the longitudinal and
transverse parts for the projection operator:
\begin{equation}
  M_{\delta\alpha}^A =  M_{\delta\alpha}^A{}_\parallel +
   M_{\delta\alpha}^A{}_\perp\,,
\label{rhomeson2}
\end{equation}
where only the longitudinal part is relevant in the present study
and is given by
 \begin{eqnarray}
M^A_\parallel &=& -i\frac{1}{4} \, \frac{m_A(\epsilon_{(\lambda)}^* n_+)}{2}
 \not\! n_- \gamma_5\,\Phi_\parallel(u)
-\frac{i m_A}{4}  \,\frac{m_A(\epsilon_{(\lambda)}^* n_+)}{2E}
 \, \Bigg\{\frac{i}{2}\,\sigma_{\mu\nu}\gamma_5 \,  n_-^\mu  n_+^\nu \,
 h_\parallel^{(t)}(u)
\nonumber\\
&& \hspace*{-0.0cm} + \,i E\int_0^u dv \,(\Phi_\perp(v) -
h_\parallel^{(t)}(v)) \
     \sigma_{\mu\nu} \gamma_5 n_-^\mu
     \, \frac{\partial}{\partial k_\perp{}_\nu}
  - \gamma_5\frac{h_\parallel'{}^{(p)}(u)}{2}\Bigg\}\, \Bigg|_{k=u
  p}\,,
\end{eqnarray}
with the momentum of the quark $q_1$ in the $A$ meson being
\begin{eqnarray}
k_1^\mu = u E n_-^\mu +k_\perp^\mu + \frac{k_\perp^2}{4
uE}n_+^\mu\,,
\end{eqnarray}
for which $E$ is the energy of the axial-vector meson and the term proportional to $k_\perp^2$ is negligible. Here, for simplicity, we introduce two light-like vectors
$n_-^\mu\equiv (1,0,0,-1)$ and $n_+^\mu\equiv (1,0,0,1)$. In
general, the QCD factorization amplitudes can be recast to the form
$\int_0^1 du \, {\rm Tr} (M^A_\parallel \cdots)$.

The LCDAs $\Phi_{\parallel,\perp}^{A}(u)$ can be expanded in terms of
Gegenbauer polynomials of the form:
\begin{eqnarray}\label{eq:t2-DA}
\Phi_{\parallel (\perp)}^{A}(u)= 6u\bar u
f_A\bigg[a_0^{\parallel\, (\perp),A}+\sum_{i=1}^\infty
a_i^{\parallel\, (\perp),A} C_i^{3/2}(2u-1)\bigg]\,.
\end{eqnarray}
where the relevant decay constants in the above equation will be
specified later. In the following we will discuss the LCDAs of
$^1P_1$ and $^3P_1$ states separately:

\subsubsection{$^1P_1$ mesons}
For the $\Phi_{\perp (\parallel)}^{^1P_1}(u)$, due to the
$G$-parity, only terms with odd (even) Gegenbauer moments survive
in the SU(3) limit. Hence, the normalization condition for the
twist-2 LCDA $\Phi_\perp^{^1P_1}$ can be chosen as
 \begin{eqnarray}
 \int_0^1 du \Phi_\perp^{^1P_1}(u) = f^\perp_{^1P_1}.
 \en
In the present work, we consider the approximation
\begin{eqnarray}
\Phi_\perp^{^1P_1} (u) =  f^\perp_{^1P_1}6 u \bar u
 \left\{ 1 +3a_1^{\perp,^1P_1}(2u-1)+a_2^{\perp,^1P_1}\, \frac{3}{2} \bigg[ 5 (2u-1)^2  - 1 \bigg]
 \right\}. \label{eq:t21P1-1}
 \en
Likewise, we take
 \be
 \Phi_\parallel^{^1P_1} (u) & = &  f^\perp_{^1P_1}6 u \bar u \left\{a_0^{\parallel,^1P_1}+3a_1^{\parallel,^1P_1}(2u-1)+a_2^{\parallel,^1P_1}\,
 \frac{3}{2} \bigg[ 5 (2u-1)^2  - 1 \bigg]\right\},
 \label{eq:t21P1-2}
\end{eqnarray}
with the normalization condition
\begin{eqnarray} \label{eq:t2nor}
 \int_0^1 du
 \Phi_\parallel^{^1P_1}(u)=f_{^1P_1}.
 \en
This normalization together with Eq. (\ref{eq:evendef1}) leads to
Eq. (\ref{eq:decayc}) for the definition of  the $^1P_1$ meson
decay constant. Eqs. (\ref{eq:t21P1-2}) and (\ref{eq:t2nor}) lead
to the relation
 \be \label{eq:f1p1}
 f_{^1P_1}=f_{^1P_1}^\perp(\mu) a_0^{\parallel,^1P_1}(\mu).
 \en
The scale dependence of $f_{^1P_1}^\perp$ must be compensated by
that of the Gegenbauer moment $a_0^{\parallel,^1P_1}$ to ensure the
scale independence of $f_{^1P_1}$. In principle, we can also use
the decay constant $f_{^1P_1}$ to construct the LCDA
$\Phi_\parallel^{^1P_1}$. However, $f_{^1P_1}$ vanishes for the
neutral $b_1(1235)$ and is very small for the charged $b_1(1235)$.
This implies a vanishing or very small $\Phi_\parallel^{^1P_1}$
unless the Gegenbauer moments $a_i^{\parallel,^1P_1}$ are very
large. Hence, it is more convenient to employ the non-vanishing
decay constant $f^\perp_{^1P_1}$ to construct
$\Phi_\parallel^{^1P_1}$. This is very similar to the scalar meson
case where the twist-2 light-cone distribution amplitude $\Phi_S$
is expressed in the form \cite{CCY}
 \be  \label{eq:Swf}
 \Phi_S(x,\mu)=\bar f_S(\mu)\,6x(1-x)\left[B_0(\mu)+\sum_{m=1}^\infty
 B_m(\mu)\,C_m^{3/2}(2x-1)\right],
 \en
with $\bar f_S$ being defined as $\la S|\bar q_2q_1|0\ra=m_S\bar
f_S$. Now $\Phi_\parallel^{^1P_1}$ can be recast to the form
 \be
 \Phi_\parallel^{^1P_1} (u) & = &  f_{^1P_1}6 u \bar u \left\{1+\mu_{^1P_1}
 \sum_{i=1}^2a_i^{\parallel,^1P_1}C_i^{3/2}(2u-1)
 \right\},
 \label{eq:t21P1-3}
\end{eqnarray}
with $\mu_{^1P_1}=1/a_0^{\parallel,^1P_1}$. For the neutral
$b_1(1235)$, $f_{b_1}$ vanishes and $\mu_{b_1}$ becomes divergent,
but the combination $f_{b_1}\mu_{b_1}$ is finite \cite{Diehl}.
Recall that for the scalar meson case, its LCDA also can be
expressed in the form
 \be \label{eq:twist2wf}
 \Phi_S(x,\mu)=f_S\,6x(1-x)\left[1+\mu_S\sum_{m=1}^\infty
 B_m(\mu)\,C_m^{3/2}(2x-1)\right],
 \en
where $\bar f_S=\mu_S f_S$ and the equation of motion leads to
$\mu_S=m_S/(m_2(\mu)-m_1(\mu))$. However, unlike the case for
scalar mesons, the decay constants $f^\perp_{^1P_1}$ and
$f_{^1P_1}$ cannot be related by equations of motion.

When the three-parton distributions and terms proportional to the
light quark masses are neglected, the twist-3 distribution
amplitudes can be related to the twist-2 $\Phi_\perp(u)$ for the
transversely polarized axial-vector meson by Wandzura-Wilczek
relations \cite{YangNP}:
\begin{eqnarray}\label{eq:ww}
 && h_\parallel^{(t)}(v)= (2v-1)\Bigg[ \int_0^v
\frac{\Phi_\perp(u)}{\bar u}du -\int_v^1
\frac{\Phi_\perp(u)}{u}du \Bigg] \equiv (2v-1) \Phi_a(v)\,, \nonumber\\
 && h_\parallel^{\prime(p)}(v)= -2\Bigg[ \int_0^v
\frac{\Phi_\perp(u)}{\bar u}du -\int_v^1
\frac{\Phi_\perp(u)}{u}du \Bigg] = -2 \Phi_a(v)\,, \nonumber\\
 &&\int_0^v du \big( \Phi_\perp (u) -h^{(t)}_\parallel
(u))= v\bar v\Bigg[ \int_0^v \frac{\Phi_\perp(u)}{\bar u}du
-\int_v^1 \frac{\Phi_\perp(u)}{u}du\Bigg] = v \bar v \Phi_a(v) \,.
\end{eqnarray}
The twist-3 LCDA $\Phi_a(u,\mu)$ satisfies the normalization
 \be \label{eq:norPhia}
 \int^1_0 \Phi_a(u)du=0,
 \en
and has the general expression
 \be
 \Phi_a^{^1P_1}(u)=3f^\perp_{^1P_1}\left[(2u-1)+\sum_{n=1}^\infty
 a_{n}^{\perp,^1P_1}(\mu)P_{n+1}(2u-1)\right],
 \en
where $P_n(u)$ are the Legendre polynomials.

\subsubsection{$^3P_1$ mesons}
In analogue to the $^1P_1$ case, we consider the approximations:
\begin{eqnarray}
\Phi_\parallel^{^3P_1} (u)& = & f_{^3P_1}6 u \bar u
 \left\{ 1 +3a_1^{\parallel,^3P_1}(2u-1)+a_2^{\parallel,^3P_1}\, \frac{3}{2} \bigg[ 5 (2u-1)^2  - 1 \bigg]
 \right\}, \label{eq:t2-1}\\
 \Phi_\perp^{^3P_1} (u) & = & f_{^3P_1}6 u \bar u\left\{a_0^{\perp,^3P_1}+3a_1^{\perp,^3P_1}(2u-1)+a_2^{\perp,^3P_1}\,
 \frac{3}{2} \bigg[ 5 (2u-1)^2  - 1 \bigg]\right\}. \label{eq:t23P1-2}
\end{eqnarray}
In the SU(3) limit, only terms with even (odd) Gegenbauer moments
for $\Phi^{^3P_1}_{\parallel(\perp)}$ survive due to the
$G$-parity. Hence, $a_1^{\parallel,^3P_1}$ and
$a_{0,2}^{\perp,^3P_1}$ vanish in the SU(3) limit. The LCDAs
respect the normalization conditions
 \begin{eqnarray}
 \int_0^1 du \Phi_\parallel^{^3P_1}(u) = f_{^3P_1}, \qquad \int_0^1 du \Phi_\perp^{^3P_1}(u)=
f_{^3P_1}^\perp,
 \en
and
\begin{equation}
 \int_0^1 du h_\parallel^{(t)}(u)
 =\int_0^1 duh_\parallel^{(p)}(u)=0.
 \end{equation}
The latter is valid in the SU(3) limit. Therefore, we obtain
 \be
 f_{^3P_1}^\perp(\mu)=f_{^3P_1}a_0^{\perp,^3P_1}(\mu),
 \en
and
 \be \label{eq:t23P1-3}
 \Phi_\perp^{^3P_1} (u) & = & f_{^3P_1}^\perp6 u \bar u\left\{1+\mu_{^3P_1}\sum_{i=1}^2a_i^{\perp,^3P_1}C_i^{3/2}(2u-1)
 \right\},
 \en
with $\mu_{^3P_1}=1/a_0^{\perp,^3P_1}$. The twist-3 LCDA $\Phi_a$
has the expression
 \be
 \Phi_a^{^3P_1}(u,\mu)=3f_{^3P_1}\left[a_0^{\perp,^3P_1}(2u-1)+\sum_{n=1}^\infty
  a_{n}^{\perp,^3P_1}(\mu)P_{n+1}(2u-1)\right].
 \en

Most of the relevant Gegenbauer moments $a_i^{\parallel\, (\perp),A}$ have been
evaluated using the QCD sum rule method \cite{YangNP}. The results
are summarized in Table \ref{tab:Gegenbauer}.

For the pseudoscalar meson LCDAs we use
\begin{eqnarray}
\Phi_P (u)& = & f_P6 u \bar u \left\{ 1 + 3 a_1^P (2u-1)  +
a_2^P\, \frac{3}{2} \bigg[ 5 (2u-1)^2  - 1 \bigg]
 \right\}, \nonumber\\
 \Phi_p (u) & = & f_P, \qquad \frac{\Phi_\sigma (u)}{6}=f_Pu(1-u),
\label{eq:pseudoscalar-DA}
\end{eqnarray}
where $\Phi_p$ and $\Phi_\sigma$ are twist-3 LCDAs. We shall
employ the sum rule results for the Gegenbauer moments of pseudoscalar
mesons \cite{Ball}
 \be
 \mu=1.0\,{\rm GeV}: && a_1^K=0.06\pm0.03\,, \quad a_2^K=0.25\pm0.15\,, \quad
 a_1^\pi=0,\quad a_2^\pi=0.25\pm0.15\,, \non \\
 \mu=2.1\,{\rm GeV}: && a_1^K=0.05\pm0.02\,, \quad a_2^K=0.17\pm0.10\,, \quad
 a_1^\pi=0,\quad a_2^\pi=0.17\pm0.10\,.
 \en
Note that in this paper the $G$-parity violating parameters
($a_1^K, a_1^{\parallel,K_{1A}}, a_{0,2}^{\perp,K_{1A}},
a_1^{\perp,K_{1B}}$ and $a_{0,2}^{\parallel,K_{1B}})$ are for
mesons containing a strange quark. For mesons involving an
anti-strange quark, the signs of $G$-parity violating parameters
have to be flipped due to the $G$-parity. The integral of the $B$
meson wave function is parameterized as \cite{BBNS}
 \begin{eqnarray}
 \int_0^1 \frac{d\rho}{1-\rho}\Phi_1^B(\rho) \equiv
 \frac{m_B}{\lambda_B}\,,
 \end{eqnarray}
where $1-\rho$ is the momentum fraction carried by the light
spectator quark in the $B$ meson. Here we use $\lambda_B ({\rm
1~GeV}) =(350\pm 100)$~MeV.

%
\begin{table}[htb!]
\caption[]{Gegenbauer moments of $\Phi_\perp$ and $\Phi_\parallel$
for $1^3P_1$ and $1^1P_1$ mesons, respectively, taken from
\cite{YangNP}.} \label{tab:Gegenbauer}
\renewcommand{\arraystretch}{1.8}
\addtolength{\arraycolsep}{0.4pt} {\small
$$
\begin{array}{|c|c|c|c|c|c|c|}\hline
 \mu & a_2^{\parallel, a_1(1260)}& a_2^{\parallel,f_1^{^3P_1}}
 & a_2^{\parallel,f_8^{^3P_1}} & a_2^{\parallel, K_{1A}}
 & \multicolumn{2}{|c|}{a_1^{\parallel, K_{1A}}}
 \\ \hline
\begin{array}{c} {\rm 1~GeV}   \\ {\rm 2.2~GeV} \end{array}&
\begin{array}{c} -0.02\pm 0.02 \\ -0.01\pm 0.01  \end{array}&
\begin{array}{c}  -0.04\pm 0.03 \\ -0.03 \pm 0.02 \end{array}&
\begin{array}{c} -0.07\pm 0.04\\ -0.05 \pm  0.03 \end{array}&
\begin{array}{c} -0.05\pm 0.03\\  -0.04 \pm 0.02 \end{array}&
\multicolumn{2}{c|}{\begin{array}{c} {0.00\pm 0.26}\\ 0.00\pm
0.22\end{array}}
\\ \hline\hline
 \mu & a_1^{\perp, a_1(1260)}& a_1^{\perp,f_1^{^3P_1}}
 & a_1^{\perp,f_8^{^3P_1}} & a_1^{\perp, K_{1A}}
 & a_0^{\perp, K_{1A}}    & a_2^{\perp, K_{1A}}
 \\ \hline
\begin{array}{c} {\rm 1~GeV}   \\ {\rm 2.2~GeV} \end{array}&
\begin{array}{c} -1.04\pm 0.34 \\ -0.83\pm 0.27  \end{array}&
\begin{array}{c} -1.06\pm 0.36 \\ -0.84\pm 0.29   \end{array}&
\begin{array}{c} -1.11\pm 0.31 \\ -0.90\pm 0.25   \end{array}&
\begin{array}{c} -1.08\pm 0.48 \\ -0.88\pm 0.39   \end{array}&
\begin{array}{c}  0.08\pm 0.09\\   0.07\pm 0.08   \end{array}&
\begin{array}{c}  0.02\pm 0.20\\   0.01\pm 0.15   \end{array}
\\ \hline\hline
\mu & a_1^{\parallel, b_1(1235)}& a_1^{\parallel,h_1^{^1P_1}}
 & a_1^{\parallel,h_8^{^1P_1}} & a_1^{\parallel, K_{1B}}
 & a_0^{\parallel, K_{1B}}    & a_2^{\parallel, K_{1B}}
 \\ \hline
\begin{array}{c} {\rm 1~GeV}   \\   {\rm 2.2~GeV} \end{array}&
\begin{array}{c} -1.95\pm 0.35 \\  -1.61\pm 0.29  \end{array}&
\begin{array}{c} -2.00\pm 0.35 \\  -1.65\pm 0.29  \end{array}&
\begin{array}{c} -1.95\pm 0.35 \\  -1.61\pm 0.29  \end{array}&
\begin{array}{c} -1.95\pm 0.45 \\  -1.57\pm 0.37  \end{array}&
\begin{array}{c}  0.14\pm 0.15 \\   0.14\pm 0.15  \end{array}&
\begin{array}{c}  0.02\pm 0.10 \\   0.01\pm 0.07  \end{array}
\\ \hline\hline
 \mu & a_2^{\perp, b_1(1235)}& a_2^{\perp,h_1^{^1P_1}}
 & a_2^{\perp,h_8^{^1P_1}} & a_2^{\perp, K_{1B}}
 & \multicolumn{2}{|c|}{a_1^{\perp, K_{1B}}}
 \\ \hline
\begin{array}{c}  {\rm 1~GeV}  \\   {\rm 2.2~GeV} \end{array}&
\begin{array}{c}  0.03\pm 0.19 \\   0.02\pm 0.15  \end{array}&
\begin{array}{c}  0.18\pm 0.22 \\  0.14 \pm 0.17  \end{array}&
\begin{array}{c}   0.14\pm 0.22\\  0.11 \pm  0.17 \end{array}&
\begin{array}{c}  -0.02\pm 0.22\\ -0.02 \pm 0.17  \end{array}&
\multicolumn{2}{c|}{\begin{array}{c}0.17\pm 0.22\\
                                    0.14\pm 0.18\end{array}}
\\ \hline
\end{array}
$$}
\end{table}
%

\subsection{Other input parameters}

For the CKM matrix elements, we use the Wolfenstein parameters
$A=0.818$, $\lambda=0.22568$, $\bar \rho=0.141$ and $\bar
\eta=0.348$ \cite{CKMfitter}. The corresponding three unitarity
angles are $\alpha=90.0^\circ$, $\beta=22.1^\circ$ and
$\gamma=68.0^\circ$.

For the running quark masses we shall use
 \be \label{eq:quarkmass}
 && m_b(m_b)=4.2\,{\rm GeV}, \qquad~~~ m_b(2.1\,{\rm GeV})=4.95\,{\rm
 GeV}, \qquad m_b(1\,{\rm GeV})=6.89\,{\rm
 GeV}, \non \\
 && m_c(m_b)=1.3\,{\rm GeV}, \qquad~~~ m_c(2.1\,{\rm GeV})=1.51\,{\rm
 GeV}, \non \\
 && m_s(2.1\,{\rm GeV})=90\,{\rm MeV}, \quad m_s(1\,{\rm GeV})=119\,{\rm
 MeV}, \non\\
 && m_d(1\,{\rm GeV})=6.3\,{\rm  MeV}, \quad~ m_u(1\,{\rm GeV})=3.5\,{\rm
 MeV}.
 \en
The uncertainty of the strange quark mass is assigned to be
$m_s(2.1\,{\rm GeV})=90\pm20$ MeV.

\section{$B\to AP$ decays in QCD factorization}

 We shall use the QCD factorization approach
\cite{BBNS,BN} to study the short-distance contributions to the
$B\to AP$ decays with $A=a_1(1260)$, $f_1(1285)$, $f_1(1420)$,
$K_1(1270)$, $b_1(1235)$, $h_1(1170)$, $h_1(1380)$, $K_1(1400)$
and $P=\pi,K$. It should be stressed that in order to define the
LCDAs of axial-vector mesons properly, it is necessary to include
the decay constants. However, for practical calculations, it is
more convenient to factor out the decay constants in the LCDAs and
put them back in the appropriate places. Recall that
$\Phi_\parallel^{^1P_1}$ has two equivalent expressions, namely,
Eqs. (\ref{eq:t21P1-2}) and (\ref{eq:t21P1-3}). However, we found
out that it is most convenient to use Eq. (\ref{eq:t21P1-2}) for
the LCDA $\Phi_\parallel^{^1P_1}$ which amounts to treating the
axial-vector decay constant of $^1P_1$ as $f_{^1P_1}^\perp$. (Of
course, this does not mean that $f_{^1P_1}$ is equal to
$f_{^1P_1}^\perp$.) Likewise, we shall use Eq. (\ref{eq:t23P1-3})
rather than Eq. (\ref{eq:t23P1-2}) for the LCDA
$\Phi_\perp^{^3P_1}$.

In QCD factorization, the factorizable amplitudes of above-mentioned
decays are collected in Appendix A. They are expressed in terms of the
flavor operators $a_i^p$ and the annihilation operators $b_i^p$ with
$p=u,c$ which can be calculated in the QCD factorization approach \cite{BBNS}.
The flavor operators $a_i^p$ are basically the Wilson coefficients in
conjunction with short-distance nonfactorizable corrections such as vertex
corrections and hard spectator interactions. In general, they have the
expressions \cite{BBNS,BN}
 \be \label{eq:ai}
  a_i^p(M_1M_2) &=&
 \left(c_i+{c_{i\pm1}\over N_c}\right)N_i(M_2) \int_0^1 \Phi_\parallel^{M_2}(x)
 dx \nonumber\\
  && + {c_{i\pm1}\over N_c}\,{C_F\alpha_s\over
 4\pi}\Big[V_i(M_2)+{4\pi^2\over N_c}H_i(M_1M_2)\Big]+P_i^p(M_2),
 \en
where $i=1,\cdots,10$,  the upper (lower) signs apply when $i$ is
odd (even), $c_i$ are the Wilson coefficients,
$C_F=(N_c^2-1)/(2N_c)$ with $N_c=3$, $M_2$ is the emitted meson
and $M_1$ shares the same spectator quark with the $B$ meson. The
quantities $V_i(M_2)$ account for vertex corrections,
$H_i(M_1M_2)$ for hard spectator interactions with a hard gluon
exchange between the emitted meson and the spectator quark of the
$B$ meson and $P_i(M_2)$ for penguin contractions. The expression
of the quantities $N_i(M_2)$ reads
 \be
 N_i(M_2)=\cases{0, & $i=6,8$~and~$M_2=A$, \cr
                 1, & else. \cr}
 \en
Note that $N_i(M_2)$ vanishes for $i=6,8$ and $M_2=A$ as a
consequence of Eq. (\ref{eq:norPhia}). The subscript $\parallel$
of $\Phi$ in the first term of Eq. (\ref{eq:ai}) reminds us of the fact that it is the
longitudinal component of the axial-vector meson's LCDA that
contributes to the $B\to AP$ decay amplitude. Specifically, we
have
 \be
 \int^1_0du\Phi_\parallel^{^1P_1}(u)=a_0^{\parallel,^1P_1}, \qquad
 \int^1_0du\Phi_\parallel^{^3P_1}(u)=1,
 \en
where we have factored out the decay constant $f_{^1P_1}^\perp$
($f_{^3P_1}$)  of $\Phi_\parallel^{^1P_1}$
($\Phi_\parallel^{^3P_1}$).

The vertex corrections in Eq. (\ref{eq:ai}) are given by
\begin{equation}\label{vertex0}
   V_i(M_{2}) = \left\{\,\,
   \begin{array}{ll}
    {\displaystyle \int_0^1\!dx\,\Phi_{M_2}(x)\,
     \Big[ 12\ln\frac{m_b}{\mu} - 18 + g(x) \Big]} \,; & \qquad
     (i=\mbox{1--4},9,10), \\[0.4cm]
   {\displaystyle \int_0^1\!dx\,\Phi_{M_2}\,
     \Big[ - 12\ln\frac{m_b}{\mu} + 6 - g(1-x) \Big]} \,; & \qquad
     (i=5,7), \\[0.4cm]
   {\displaystyle \int_0^1\!dx\, \Phi_{m_2}(x)\,\Big[ -6 + h(x) \Big]}
    \,; & \qquad (i=6,8),
   \end{array}\right.
\end{equation}
with
 \begin{eqnarray}
 g(x)&=& 3\Bigg( \frac{1-2x}{1-x} \ln x -i\pi\Bigg)\nonumber\\
  && + \Big[ 2{\rm Li}_2(x) -\ln^2x +\frac{2\ln x}{1-x} -(3+2i\pi)\ln x -
 (x \leftrightarrow 1-x)\Big]\,, \nonumber\\
 h(x)&=&  2{\rm Li}_2(x) -\ln^2x -(1+2i\pi)\ln x -
 (x \leftrightarrow 1-x) \,,
 \end{eqnarray}
where $\Phi_M$ ($\Phi_m$) is the twist-2 (twist-3) light-cone
distribution amplitude of the meson $M$. More specifically,
$\Phi_M=\Phi_P$, $\Phi_m=\Phi_p$ for $M=P$ and
$\Phi_M=\Phi_\parallel^A$, $\Phi_m=\Phi_a$ for $M=A$. For the
general LCDAs
 \be
 \Phi_M(x) &=& 6x(1-x)\left[\alpha_0+\sum_{n=1}^\infty
 \alpha_n(\mu)C_n^{3/2}(2x-1)\right], \non \\
 \Phi_m(x) &=& \beta_0+\sum_{n=1}^\infty
 \beta_n(\mu)P_n(2x-1),
 \en
the vertex corrections read
 \be
 V_i(M)= \left(12\ln{m_b\over\mu}-18-{1\over
 2}-3i\pi\right)\alpha_0+\left({11\over
 2}-3i\pi\right)\alpha_1-{21\over 20}\alpha_2+\left({79\over 36}-{2i\pi\over
 3}\right)\alpha_3+\cdots,
 \en
for~$i=1-4,9,10,$
 \be
 V_i(M)=\left(-12\ln{m_b\over\mu}+6+{1\over
 2}+3i\pi\right)\alpha_0+\left({11\over
 2}-3i\pi\right)\alpha_1+{21\over 20}\alpha_2+\left({79\over 36}+{2i\pi\over
 3}\right)\alpha_3+\cdots,
  \en
for~$i=5,7,$
 \be
 V_i(M)=
 -6\beta_0+(3-i2\pi)\beta_1+\left({19\over 18}-{i\pi\over
 3}\right)\beta_3+\cdots,
 \en
for~$i=6,8.$

As for the hard spectator function $H$, it has the expression
 \be \label{eq:Hi}
 H_i(M_1M_2) &=& {-if_B f_{M_1} f_{M_2} \over X^{(\overline{B} M_1, M_2)}}
 \int^1_0 {d\rho\over\rho}\, \Phi_B(\rho)\int^1_0 {d\xi\over
\bar\xi} \,\Phi_{M_2}(\xi)\int^1_0 {d\eta\over \deta}\left[
\Phi_{M_1}(\eta)\pm r_\chi^{M_1}\,{\bar\xi\over
\xi}\,\Phi_{m_1}(\eta)\right], \non \\
 \en
for $i=1-4,9,10$, where the upper sign is for $M_1=P$ and the
lower sign for $M_1=A$,
 \be \label{eq:Hi57}
 H_i(M_1M_2) &=& {if_B f_{M_1} f_{M_2} \over X^{(\overline{B} M_1, M_2)}}\int^1_0 {d\rho\over\rho}\,
\Phi_B(\rho)\int^1_0 {d\xi\over \xi} \,\Phi_{M_2}(\xi)\int^1_0
{d\eta\over \deta}\left[ \Phi_{M_1}(\eta)\pm
r_\chi^{M_1}\,{\xi\over
\bar\xi}\,\Phi_{m_1}(\eta)\right], \non \\
 \en
for $i=5,7$ and $H_i=0$ for $i=6,8$,  $\bar\xi\equiv 1-\xi$ and
$\bar\eta\equiv 1-\eta$.  In above equations,
 \be \label{eq:rchi}
 r_\chi^P={2m_P^2\over m_b(\mu)(m_2+m_1)(\mu)}, \qquad
 r_\chi^A={2m_A\over m_b(\mu)}\,,
 \en
and  $X^{(\ov B M_1, M_2)}$ is the factorizable amplitude defined
in Eq. (\ref{eq:X}).

Weak annihilation contributions are described by the terms $b_i$,
and $b_{i,{\rm EW}}$ in Eq. (\ref{eq:beta}) which have the
expressions
 \be \label{eq:bi}
 b_1 &=& {C_F\over N_c^2}c_1A_1^i, \qquad\quad b_3={C_F\over
 N_c^2}\left[c_3A_1^i+c_5(A_3^i+A_3^f)+N_cc_6A_3^f\right], \non \\
 b_2 &=& {C_F\over N_c^2}c_2A_1^i, \qquad\quad b_4={C_F\over
 N_c^2}\left[c_4A_1^i+c_6A_2^f\right], \non \\
 b_{\rm 3,EW} &=& {C_F\over
 N_c^2}\left[c_9A_1^{i}+c_7(A_3^{i}+A_3^{f})+N_cc_8A_3^{i}\right],
 \non \\
 b_{\rm 4,EW} &=& {C_F\over
 N_c^2}\left[c_{10}A_1^{i}+c_8A_2^{i}\right],
 \en
where the subscripts 1,2,3 of $A_n^{i,f}$ denote the annihilation
amplitudes induced from $(V-A)(V-A)$, $(V-A)(V+A)$ and
$(S-P)(S+P)$ operators, respectively, and the superscripts $i$ and
$f$ refer to gluon emission from the initial and final-state
quarks, respectively. Their explicit expressions are:
 \be \label{eq:ann}
 A_1^{i}&=& \int\cdots \cases{
 \left( \Phi_P(x)\Phi_A(y)\left[{1\over y(1-x\bar y)}+{1\over
 \bar x^2y}\right]-r_\chi^Ar_\chi^P\Phi_p(x)\Phi_a(y)\,{2\over
 \bar xy}\right); & for~$M_1M_2=AP$,  \cr
 \left( \Phi_A(x)\Phi_P(y)\left[{1\over y(1-x\bar y)}+{1\over
 \bar x^2y}\right]-r_\chi^Ar_\chi^P\Phi_a(x)\Phi_p(y)\,{2\over
 \bar xy}\right); & for~$M_1M_2=PA$, } \non  \\
 A_2^{i}&=& \int\cdots \cases{
 \left(\Phi_P(x)\Phi_A(y)\left[{1\over \bar x(1-x\bar y)}+{1\over
 \bar xy^2}\right]-r_\chi^Ar_\chi^P\Phi_p(x)\Phi_a(y)\,{2\over
 \bar xy}\right); & for~$M_1M_2=AP$,  \cr
 \left( \Phi_A(x)\Phi_P(y)\left[{1\over \bar x(1-x\bar y)}+{1\over
 \bar xy^2}\right]-r_\chi^Ar_\chi^P\Phi_a(x)\Phi_p(y)\,{2\over
 \bar xy}\right); & for~$M_1M_2=PA$, }  \non \\
 A_3^{i}&=& \int\cdots \cases{ \left( -r_\chi^A\Phi_P(x)\Phi_a(y)
 \,{2\bar y\over \bar xy(1-x\bar y)}-r_\chi^P\Phi_p(x)\Phi_A
 (y)\,{2x\over \bar xy(1-x\bar y)}\right); & for~$M_1M_2=AP$, \cr
 \left( r_\chi^P\Phi_A(x)\Phi_p(y)
 \,{2\bar y\over \bar xy(1-x\bar y)}+r_\chi^A\Phi_a(x)\Phi_P
 (y)\,{2x\over \bar xy(1-x\bar y)}\right); & for~$M_1M_2=PA$,}
 \non \\
 A_3^{f} &=& \int\cdots \cases{ \left(-r_\chi^A
 \Phi_P(x)\Phi_a(y)\,{2(1+\bar x)\over \bar x^2y}+r_\chi^P
 \Phi_p(x)\Phi_A(y)\,{2(1+y)\over \bar xy^2}\right); &
 for~$M_1M_2=AP$, \cr
 \left( r_\chi^P
 \Phi_A(x)\Phi_p(y)\,{2(1+\bar x)\over \bar x^2y}-r_\chi^A
 \Phi_a(x)\Phi_P(y)\,{2(1+y)\over \bar xy^2}\right); &
 for~$M_1M_2=PA$,} \non \\
  A_1^f &=& A_2^f=0,
 \en
where $\int\cdots=\pi\alpha_s \int^1_0dxdy$, $\bar x=1-x$ and
$\bar y=1-y$. Note that we have adopted the same convention as in
\cite{BN} that $M_1$ contains an antiquark from the weak vertex
with longitudinal fraction $\bar y$, while $M_2$ contains a quark
from the weak vertex with momentum fraction $x$.

Using the asymptotic distribution amplitudes for
$\Phi_P,\Phi_p,\Phi_\parallel^{^3P_1},\Phi_a^{^1P_1}$ and the
leading contributions to $\Phi_\parallel^{^1P_1},\Phi_a^{^3P_1}$:
 \be
&& \Phi_P(u)=6u\bar u, \qquad \Phi_\parallel^{^3P_1}(u)=6u\bar u,
 \qquad \Phi_\parallel^{^1P_1}(u)=18\,a_1^{\parallel,^1P_1}u\bar
 u(2u-1), \non \\
&& \Phi_p(u)=1, \qquad
\Phi_a^{^3P_1}(u)=3\,a_1^{\perp,^3P_1}(6u^2-6u+1),
 \qquad \Phi_a^{^1P_1}(u)=3(2u-1),
 \en
we obtain from Eq. (\ref{eq:ann}) that
\begin{eqnarray}
 A_1^i(^3P_1\,P) &\approx &  6\pi\alpha_s \left[\,
    3\,\bigg( X_A - 4 + \frac{\pi^2}{3} \bigg)
    - a_1^{\perp,^3P_1} r_\chi^{^3P_1} r_\chi^{P} X_A (X_A -3) \right] ,
    \non \\
 A_1^i(^1P_1\,P) &\approx &  6\pi\alpha_s
 \left[\,-3a_1^{\parallel,^1P_1} (X_A+29-3\pi^2)+ r_\chi^{^1P_1} r_\chi^{P} X_A (X_A -2)\right], \non \\
 A_2^i(^3P_1\,P) &\approx &  6\pi\alpha_s \left[\,
    3\,\bigg( X_A - 4 + \frac{\pi^2}{3} \bigg)
    - a_1^{\perp,^3P_1} r_\chi^{^1P_1} r_\chi^{P} X_A (X_A -3) \right] ,  \non \\
 A_2^i(^1P_1\,P) &\approx &  6\pi\alpha_s
 \left[\,-3a_1^{\parallel,^1P_1}(3X_A+4-\pi^2)+ r_\chi^{^1P_1} r_\chi^{P} X_A (X_A -2)\right], \non \\
 A_3^i(^3P_1\,P) &\approx & 6\pi\alpha_s\, \Bigg[ -r_\chi^{P}
    \bigg( X_A^2 - 2 X_A + {\pi^2\over 3} \bigg)
    - 3a_1^{\perp,^3P_1} r_\chi^{^3P_1} \Bigg(X_A^2 - 2X_A -6 +\frac{\pi^2}{3}\Bigg)
    \Bigg] \,, \non \\
 A_3^i(^1P_1\,P) &\approx &  6\pi\alpha_s\, \Bigg[ 3a_1^{\parallel,^1P_1}r_\chi^{P}
    \bigg( X_A^2 - 4 X_A +4+ {\pi^2\over 3} \bigg)
    + 3r_\chi^{^1P_1} \Bigg(X_A^2 - 2X_A +4 -{\pi^2\over 3}\Bigg)
    \Bigg] \,, \non \\
 A_3^{f}(^3P_1\,P) & \approx& 6 \pi\alpha_s(2X_A-1) \bigg[ r_\chi^P X_A -
 3 a_1^{\perp, ^3P_1} r_\chi^{^3P_1} (X_A-3) \bigg]\,, \non
 \\
 A_3^{f}(^1P_1\,P) & \approx& 6 \pi\alpha_s \bigg[ -a_1^{\parallel,^1P_1}r_\chi^P X_A(6X_A-11)
 + 3 r_\chi^{^1P_1} (2X_A-1)(X_A-2) \bigg]\,,
 \end{eqnarray}
and
 \be
&& A_1^i(P\,^3P_1)=A_1^i(^3P_1\,P), \qquad~~
A_1^i(P\,^1P_1)=-A_2^i(^1P_1\,P), \non \\
&& A_2^i(P\,^3P_1)=A_2^i(^3P_1\,P), \qquad~~
A_2^i(P\,^1P_1)=-A_1^i(^1P_1\,P), \non \\
&& A_3^i(P\,^3P_1)=-A_3^i(^3P_1\,P), \qquad
A_3^i(P\,^1P_1)=A_3^i(^1P_1\,P), \non \\
&& A_3^f(P\,^3P_1)=A_3^f(^3P_1\,P), \qquad~~
A_3^f(P\,^1P_1)=-A_3^f(^1P_1\,P),
 \en
where the logarithmic divergences occurred in weak annihilation
are described by the variable $X_A$
 \be
 \int_0^1 {du\over u}\to X_A, \qquad \int_0^1 {\ln u\over u}\to -{1\over 2}X_A.
 \en
Following \cite{BBNS}, these variables are parameterized as
 \be \label{eq:XA}
 X_A=\ln\left({m_B\over \Lambda_h}\right)(1+\rho_A
 e^{i\phi_A}),
 \en
with the unknown real parameters $\rho_A$ and $\phi_A$. Likewise,
the endpoint divergence $X_H$ in the hard spectator contributions
can be parameterized in a similar manner. Following
\cite{CCY,CCYSV}, we adopt $\rho_{A,H}\leq 0.5$ and arbitrary
strong phases $\phi_{A,H}$ with $\rho_{A,H}=0$ by default.

Besides the penguin and annihilation contributions formally of
order $1/m_b$, there may exist other power corrections which
unfortunately cannot be studied in a systematical way as they are
nonperturbative in nature. The so-called ``charming penguin"
contribution is one of the long-distance effects that have been
widely discussed. The importance of this nonpertrubative effect
has also been conjectured to be justified in the context of
soft-collinear effective theory \cite{Bauer}. More recently, it
has been shown that such an effect can be incorporated in
final-state interactions \cite{CCSfsi}. However, in order to see
the relevance of the charming penguin effect to $B$ decays into
scalar resonances, we need to await more data with better
accuracy.

\section{Numerical Results}

\subsection{Branching ratios}
The calculated branching ratios for the decays $B\to A\pi,~AK$
with
$A=a_1(1260),b_1(1235),K_1(1270),K_1(1400),f_1(1285),f_1(1420),h_1(1170),h_1(1380)$
are collected in Tables \ref{tab:BRa1}-\ref{tab:BRf1}. For $B\to
A$ transition form factors we use those obtained by the sum rule
approach, Table \ref{tab:SRFF}. The theoretical errors correspond
to the uncertainties due to variation of (i) the Gegenbauer
moments (Table \ref{tab:Gegenbauer}), the axial-vector meson decay
constants, (ii) the heavy-to-light form factors and the strange
quark mass, and (iii) the wave function of the $B$ meson
characterized by the parameter $\lambda_B$, the power corrections
due to weak annihilation and hard spectator interactions described
by the parameters $\rho_{A,H}$, $\phi_{A,H}$, respectively. To
obtain the errors shown in Tables \ref{tab:BRa1}-\ref{tab:BRf1},
we first scan randomly the points in the allowed ranges of the
above seven parameters in three separated groups: the first two,
the second two and the last three, and then add errors in each
group in quadrature.

\begin{table}[th]
\caption{Branching ratios (in units of $10^{-6}$) for the decays
$B\to a_1(1260)\, \pi$, $a_1(1260)\, K$, $b_1(1235)\pi$ and
$b_1(1235)K$. The theoretical errors correspond to the
uncertainties due to variation of (i) Gegenbauer moments, decay
constants, (ii) quark masses, form factors, and (iii) $\lambda_B,
\rho_{A,H}$, $\phi_{A,H}$, respectively. Other model predictions
are also presented here for comparison. In \cite{Nardulli07},
predictions are obtained for two different sets of form factors,
denoted by I and II, respectively, corresponding to the mixing
angles $\theta_{K_1}=32^\circ$ and $58^\circ$ (see the text for
more details).}
 \label{tab:BRa1}
\begin{ruledtabular}
\begin{tabular}{l r r r r r}
Mode & CMV \cite{Calderon} & LNP(I)\cite{Nardulli07} & LNP(II) &
This work & Expt.
\cite{BaBara1pi,Bellea1pi,BaBara1pitime,BaBarb1pi,BaBara1pi07,BaBara1K,Brown}  \\
\hline
 $\overline B^0\to a_1^+ \pi^-$  & 74.3 & 4.7 & 11.8  &
 $9.1^{+0.2+2.2+1.7}_{-0.2-1.8-1.1}$ &  $12.2\pm4.5$ \footnotemark[1] \non \\
 $\overline B^0\to a_1^-\pi^+$  & 36.7 & 11.1 & 12.3 &
 $23.4^{+2.3+6.2+1.9}_{-2.2-5.5-1.3}$ & $21.0\pm5.4$ \footnotemark[1]
 \non \\
 $  \overline B^0\to a_1^\pm \pi^\mp$  & 111.0 & 15.8 & 24.1 &
 $32.5^{+2.5+8.4+3.6}_{-2.4-7.3-2.4}$ & $31.7\pm3.7$ \footnotemark[2]
 \non \\
 $B^-\to a_1^0 \pi^-$  & 43.2 & 3.9 & 8.8 & $7.6^{+0.3+1.7+1.4}_{-0.3-1.3-1.0}$ &  $20.4\pm4.7\pm3.4$  \nonumber \\
 $  \overline B^0\to a_1^0 \pi^0$  & 0.27 & 1.1 & 1.7 &
 $0.9^{+0.1+0.3+0.7}_{-0.1-0.2-0.3}$ &  \non \\
 $B^-\to a_1^- \pi^0$   & 13.6 & 4.8 & 10.6 & $14.4^{+1.4+3.5+2.1}_{-1.3-3.2-1.9}$ & $26.4\pm5.4\pm4.1$
 \non \\
 $\overline B^0\to a_1^+ K^-$  & 72.2 & 1.6 & 4.1 &
 $18.3^{+1.0+14.2+21.1}_{-1.0-~7.2-~7.5}$  &  $16.3\pm2.9\pm2.3$ \non \\
 $\overline B^0\to a_1^0 \overline K^0$ & 42.3 & 0.5 & 2.5  & $6.9^{+0.3+6.1+9.5}_{-0.3-2.9-3.2}$ & \nonumber \\
 $B^-\to a_1^- \overline K^0$ & 84.1 & 2.0 & 5.2 & $21.6^{+1.2+16.5+23.6}_{-1.1-~8.5-11.9}$
 & $34.9\pm 5.0\pm4.4$ \nonumber \\
 $B^-\to a_1^0  K^-$ & 43.4 & 1.4 & 2.8 & $13.9^{+0.9+9.5+12.9}_{-0.9-5.1-~4.9}$  \nonumber\\
 \hline
 $\overline B^0\to b_1^+ \pi^-$  & 36.2 & 6.9 & 0.7 &
 $11.2^{+0.3+2.8+2.2}_{-0.3-2.4-1.9}$  \non \\
 $\overline B^0\to b_1^-\pi^+$   &4.4 & $\approx0$  & $\approx0$ &
 $0.3^{+0.1+0.1+0.3}_{-0.0-0.1-0.1}$ \non \\
 $\overline B^0\to b_1^\pm \pi^\mp$   & 40.6 & 6.9 & 0.7 &
 $11.4^{+0.4+2.9+2.5}_{-0.3-2.5-2.0}$ & $10.9\pm1.2\pm0.9$    \non \\
 $\overline B^0\to b_1^0 \pi^0$   & 0.15 & 0.5 & 0.01 &
 $1.1^{+0.2+0.1+0.2}_{-0.2-0.1-0.2}$ \non \\
 $B^-\to b_1^- \pi^0$   & 4.2 & 4.8 & 0.5 & $0.4^{+0.0+0.2+0.4}_{-0.0-0.1-0.2}$ \non \\
 $B^-\to b_1^0 \pi^-$ & 18.6 & 4.5 & 0.4 & $9.6^{+0.3+1.6+2.5}_{-0.3-1.6-1.5}$ & $6.7\pm1.7\pm1.0$  \nonumber \\
 $\overline B^0\to b_1^+ K^-$   & 35.7 & 2.4 & 0.2  & $12.1^{+1.0+9.7+12.3}_{-0.9-4.9-30.2}$ & $7.4\pm1.0\pm1.0$ \non \\
 $\overline B^0\to b_1^0 \overline K^0$  & 19.3 & 4.1 & 0.4 & $7.3^{+0.5+5.4+6.7}_{-0.5-2.8-6.5}$ \nonumber \\
 $B^-\to b_1^- \overline K^0$  & 41.5 & 3.0 & 0.3 &
 $14.0^{+1.3+11.5+13.9}_{-1.2-~5.9-~8.3}$ & \nonumber \\
 $B^-\to b_1^0  K^-$ & 18.1 & 2.6 & 0.07 & $6.2^{+0.5+5.0+6.4}_{-0.5-2.5-5.2}$ & $9.1\pm1.7\pm1.0$
 \\
\end{tabular}
 \footnotetext[1]{BaBar data only \cite{BaBara1pitime}.}
 \footnotetext[2]{the average of BaBar \cite{BaBara1pi} and Belle \cite{Bellea1pi} data.}
\end{ruledtabular}
\end{table}

\subsubsection{$\ov B\to a_1\pi,a_1\ov K$ decays}

From Table \ref{tab:BRa1} we see that the predictions for $\ov
B^0\to a_1^\pm\pi^\mp$ are in excellent agreement with the average
of the BaBar and Belle measurements \cite{BaBara1pi,Bellea1pi}.
BaBar has also measured time-dependent \CP asymmetries in the
decays $\ov B^0\to a_1^\pm\pi^\mp$ \cite{BaBara1pitime}. Using the
measured parameter $\Delta C$ (see Sec. IV.B), BaBar is able to
determine the rates of $\ov B^0\to a_1^+\pi^-$ and $\ov B^0\to
a_1^-\pi^+$ separately, as shown in Table \ref{tab:BRa1}. It is
expected that the latter governed by the decay constant of
$a_1(1260)$ has a rate larger than the former as $f_{a_1}\gg
f_\pi$. Again, theory is consistent the data within errors.
However, there are some discrepancies between theory and
experiment for $a_1^0\pi^-$ and $a_1^-\pi^0$ modes. It appears
that the relations $\B(B^-\to a_1^0\pi^-)\gsim \B(\ov B^0\to
a_1^+\pi^-)$ and $\B(B^-\to a_1^-\pi^0)\gsim \B(\ov B^0\to
a_1^-\pi^+)$ observed by BaBar are opposite to the naive
expectation that $\B(B^-\to a_1^0\pi^-)< \B(\ov B^0\to
a_1^+\pi^-)$ and $\B(B^-\to a_1^-\pi^0)< \B(\ov B^0\to
a_1^-\pi^+)$. As for $B\to a_1K$ decays, although the agreement
with the data for $\ov B^0\to a_1^+K^-$ is excellent, the
expectation of $\B(B^-\to a_1^-\ov K^0)\sim \B(\ov B^0\to
a_1^+K^-)$ is not consistent with experiment. More specifically,
in QCDF we obtain the ratios
 \be
 R_1\equiv {\B(B^-\to a_1^0\pi^-)\over \B(\ov B^0\to
 a_1^+\pi^-)} &=& 0.83^{+0.02+0.06+0.12}_{-0.02-0.05-0.17} \qquad
 ({\rm expt:}~1.67\pm0.78), \non
 \\
 R_2\equiv {\B(B^-\to a_1^-\pi^0)\over \B(\ov B^0\to
 a_1^-\pi^+)} &=& 0.62^{+0.01+0.01+0.08}_{-0.01-0.01-0.10} \qquad
 ({\rm expt:} ~1.26\pm0.46), \non
 \\
 R_3\equiv {\B(B^-\to a_1^-\ov K^0)\over \B(\ov B^0\to
 a_1^+K^-)} &=& 1.18^{+0.01+0.02+0.04}_{-0.01-0.02-0.04} \qquad
 ({\rm expt:}~ 2.14\pm0.63).
 \en
In above ratios the hadronic uncertainties  are mainly governed by
weak annihilation and spectator scattering in $R_1$, $R_2$ and
largely canceled out in $R_3$. It is evident that while the
predicted $R_1$ is barely consistent with the data within errors,
theory does not agree with experiment for $R_2$ and $R_3$. This
should be clarified by the improved measurements of these modes in
the future.

While the tree-dominated $a_1\pi$ modes have similar rates as
$\rho\pi$ ones, the penguin-dominated $a_1K$ modes resemble much more to
$\pi K$ than $\rho K$, as first pointed out in \cite{Yanga1}.
One can see from Eqs. (\ref{eq:a1K}) and (\ref{eq:alpha}) that the
dominant penguin coefficients $\alpha_4^p(a_1K)$ and
$\alpha_4^p(\pi K)$ are constructive in $a_4^p$ and $a_6^p$
penguin coefficients:
 \be
 \alpha_4^p(a_1K)=a_4^p(a_1K)+r_\chi^K a_6^p(a_1K), \qquad
  \alpha_4^p(\pi K)=a_4^p(\pi K)+r_\chi^K a_6^p(\pi K),
 \en
whereas $\alpha_4^p(\rho K)=a_4^p(\rho K)-r_\chi^K a_6^p(\rho K)$
\cite{BN}. Consequently, when the weak annihilation contribution
is small, $B\to a_1K$ and $B\to \pi K$ decays should have similar
rates. However, if weak annihilation is important, then it will
contribute more to the $a_1K$ mode than the $\pi K$ one due to the
fact that $f_{a_1}\gg f_\pi$, recalling that the weak annihilation
amplitude is proportional to $f_Bf_{M_1}f_{M_2}b_i$. By comparing
Table \ref{tab:BRa1} of the present work with Table 2 of
\cite{BN}, we see that the default results for the branching
ratios of $B\to a_1K$ and $B\to \pi K$ decays are indeed similar,
while the hadronic uncertainties arising from weak annihilation
are bigger in the former.

\subsubsection{$\ov B\to b_1\pi,b_1\ov K$ decays}

As for $\ov B\to b_1(1235)\pi$ decays, there is a good agreement
between theory and experiment. Notice that it is naively expected
that the $b_1^-\pi^+$ mode is highly suppressed relative to the
$b_1^+\pi^-$ one as the decay amplitude of the former has the form
$a_1F_1^{B\pi}*f_{b_1}\Phi^{b_1}_\parallel$ ($a_1$ being the
effective Wilson coefficient for the color-allowed tree amplitude)
and the decay constant $f_{b_1}$ vanishes in the isospin limit. As
noted in passing, the LCDA $\Phi_\parallel^{b_1}(u)$ given by
(\ref{eq:t21P1-3}) is finite even if $f_{b_1}=0$. This is because
the coefficient $\mu_{b_1}=1/a_0^{\parallel,b_1}$ in the wave
function of $b_1$ will become divergent if $f_{b_1}=0$, but the
combination $f_{b_1}\mu_{b_1}$ is finite. More precisely,
$f_{b_1}\mu_{b_1}$ is equal to $f_{b_1}^\perp$, the transverse decay
constant of the $b_1$ [cf. Eq. (\ref{eq:f1p1})]. Therefore,
$\Phi_\parallel^{b_1}(u)$ can be recast to the form of Eq.
(\ref{eq:t21P1-2}) which amounts to replacing $f_{b_1}$ by
$f_{b_1}^\perp$ in the calculation. Now, one may wonder how to see
the suppression of $b_1^-\pi^+$ relative to $b_1^+\pi^-$ ? The key
point is the term $\int \Phi_\parallel^M(x)dx$ appearing in the
expression for the effective parameter $a_i$ [see Eq.
(\ref{eq:ai})]. This term vanishes for the $b_1$ meson in the
isospin limit. As a result, the parameter $a_1$ for the decay $\ov
B^0\to b_1^-\pi^+$ vanishes in the absence of vertex, penguin and
spectator corrections. On the contrary, $a_1=c_1+{c_2\over
3}+\cdots$ for the channel $\ov B^0\to b_1^+\pi^-$. This explains
the suppression of $\ov B^0\to b_1^-\pi^+$ relative to $b_1^+\pi^-$.
After all, the $b_1^-\pi^+$ mode does not evade the decay constant
suppression. It does receive contributions from vertex and hard
spectator corrections and weak annihilation, but they are all
suppressed. The BaBar measurement of charge-flavor asymmetry $\Delta
C$ implies the ratio $\Gamma(\ov B^0\to b_1^-\pi^+)/\Gamma(\ov
B^0\to b_1^\pm\pi^\mp)=-0.01\pm0.12$ \cite{BaBarb1pi}. This confirms
the expected suppression.

\begin{table}[t]
\caption{Same as Table \ref{tab:BRa1} except for  the decays $B\to
K_1(1270)\, \pi$, $K_1(1270)\, K$, $K_1(1400)\pi$ and $K_1(1400)K$
for two different mixing angles $\theta_{K_1}=-37^\circ$ and
$-58^\circ$ (in parentheses). In the framework of \cite{Nardulli07},
only the $K_1^-(1400)\pi^0$ and $\ov K_1^0(1400)\pi^0$ modes depend
on the mixing angle $\theta_{K_1}$. Note that the results of
\cite{Nardulli07} and \cite{Calderon} shown in the table are
obtained for $\theta_{K_1}=32^\circ$ and $58^\circ$ (in
parentheses).}
 \label{tab:BRK1}
\begin{ruledtabular}
\begin{tabular}{l l l l r}
Mode & \cite{Calderon} & \cite{Nardulli07} & This work & Expt. \cite{Blanc} \\
\hline
 $\overline B^0\to K_1^-(1270) \pi^+$  & 4.3~(4.3) & 7.6  &
 $3.0^{+0.8+1.5+4.2}_{-0.6-0.9-1.4}~(2.7^{+0.6+1.3+4.4}_{-0.5-0.8-1.5})$ &  $12.0\pm3.1^{+9.3}_{-4.5}<25.2$  \\
 $\overline B^0\to \ov K_1^0(1270) \pi^0$ & 2.3~(2.1) & 0.4 &
 $1.0^{+0.0+0.6+1.7}_{-0.0-0.3-0.6}~(0.8^{+0.1+0.5+1.7}_{-0.1-0.3-0.6})$ & \\
 $B^-\to \ov K_1^0(1270) \pi^-$ & 4.7~(4.7) & 5.8 &
 $3.5^{+0.1+1.8+5.1}_{-0.1-1.1-1.9}~(3.0^{+0.2+0.1+2.7}_{-0.2-0.2-2.2})$ &  \\
 $B^-\to K_1^-(1270) \pi^0$ & 2.5~(1.6) & 4.9 &
 $2.7^{+0.1+1.1+3.1}_{-0.1-0.7-1.0}~(2.5^{+0.1+1.0+3.2}_{-0.1-0.7-1.0})$ &  \\
 $\overline B^0\to K_1^-(1400) \pi^+$  & 2.3~(2.3) & 4.0  &
 $5.4^{+1.1+1.7+9.9}_{-1.0-1.3-2.8}~(2.2^{+1.1+0.7+2.6}_{-0.8-0.6-1.3})$  &  $16.7\pm2.6^{+3.5}_{-5.0}<21.8$  \\
 $\overline B^0\to \ov K_1^0(1400) \pi^0$ & 1.7~(1.6) & 3.0~(1.7) &
 $2.9^{+0.3+0.7+5.5}_{-0.3-0.6-1.7}~(1.5^{+0.4+0.3+1.7}_{-0.3-0.3-0.9})$ & \\
 $B^-\to \ov K_1^0(1400) \pi^-$ & 2.5~(2.5) & 3.0 &
 $6.5^{+1.0+2.0+11.6}_{-0.9-1.6-~3.6}~(2.8^{+1.0+0.9+3.0}_{-0.8-0.9-1.7})$ &  \\
 $B^-\to K_1^-(1400) \pi^0$ & 0.7~(0.6) & 1.0~(1.4) &
 $3.0^{+0.4+1.1+5.2}_{-0.4-0.7-1.3}~(1.0^{+0.4+0.4+1.2}_{-0.3-0.4-0.5})$ &  \\
 \hline
 $\overline B^0\to K_1^-(1270) K^+$  &  &   &
 $0.01^{+0.01+0.00+0.02}_{-0.00-0.00-0.01}~(0.01^{+0.00+0.00+0.02}_{-0.00-0.00-0.01})$  &   \\
 $\overline B^0\to K_1^+(1270) K^-$  &  &   &
 $0.06^{+0.01+0.00+0.46}_{-0.01-0.00-0.06}~(0.04^{+0.01+0.00+0.27}_{-0.01-0.00-0.04})$  &   \\
 $B^-\to K_1^0(1270) K^-$ & 0.22~(0.22) &  &
 $0.25^{+0.01+0.15+0.39}_{-0.01-0.08-0.09}~(0.22^{+0.01+0.12+0.39}_{-0.01-0.07-0.12})$ &  \\
 $B^-\to K_1^-(1270) K^0$ & 0.02~(0.75) &  &
 $0.05^{+0.02+0.07+0.10}_{-0.02-0.03-0.04}~(0.05^{+0.02+0.09+0.10}_{-0.01-0.03-0.04})$ &  \\
 $\overline B^0\to \ov K_1^0(1270) K^0$ & 0.02~(0.70) & &
 $2.30^{+0.16+1.13+1.43}_{-0.15-0.61-0.61}~(2.10^{+0.13+1.23+1.31}_{-0.13-0.65-0.57})$ & \\
 $\overline B^0\to K_1^0(1270) \ov K^0$ & 0.20~(0.20) & &
 $0.24^{+0.01+0.11+0.33}_{-0.01-0.07-0.13}~(0.26^{+0.10+0.12+0.47}_{-0.01-0.08-0.17})$ & \\
 $\overline B^0\to K_1^-(1400) K^+$  &  &   &
 $0.09^{+0.01+0.00+0.23}_{-0.01-0.00-0.09}~(0.07^{+0.02+0.00+0.16}_{-0.02-0.00-0.06})$  &   \\
 $\overline B^0\to K_1^+(1400) K^-$  &  &   &
 $0.02^{+0.00+0.00+0.04}_{-0.00-0.00-0.00}~(0.01^{+0.00+0.00+0.16}_{-0.00-0.00-0.00})$  &   \\
 $B^-\to K_1^0(1400) K^-$ & 0.12~(0.12) &  &
 $0.48^{+0.08+0.15+0.81}_{-0.08-0.12-0.26}~(0.22^{+0.07+0.07+0.24}_{-0.07-0.07-0.13})$ &  \\
 $B^-\to K_1^-(1400) K^0$ & 4.4~(3.9) &  &
 $0.01^{+0.00+0.01+0.14}_{-0.00-0.00-0.01}~(0.01^{+0+0.02+0.04}_{-0-0.00-0.00})$ &  \\
 $\overline B^0\to \ov K_1^0(1400) K^0$ & 4.1~(3.6) & &
 $0.08^{+0.01+0.17+0.59}_{-0.01-0.06-0.08}~(0.10^{+0.02+0.21+0.15}_{-0.02-0.08-0.10})$ & \\
 $\overline B^0\to K_1^0(1400) \ov K^0$ & 0.11~(0.11) & &
 $0.50^{+0.08+0.13+0.92}_{-0.07-0.11-0.32}~(0.25^{+0.07+0.08+0.31}_{-0.07-0.07-0.15})$ & \\
\end{tabular}
\end{ruledtabular}
\end{table}

Since $\ov B\to b_1 \ov K$ decays receive sizable annihilation
contributions, their rates are sensitive to the interference
between penguin and annihilation terms. As a consequence, the
measured branching ratios of $\ov B\to b_1 \ov K$ would provide useful
information on the sign of the $B\to b_1(1235)$ transition form
factors. We found that if the form factor $V_0^{Bb_1}$ is of the
same sign as $V_0^{Ba_1}$, $\B(\ov B\to b_1 \ov K)$ will be enhanced by
a factor of $2\sim 3$, for example, $\B(\ov B^0\to
b_1^+K^-)\approx 21\times 10^{-6}$ which is too large compared to
the experimental value of $(7.4\pm1.4)\times 10^{-6}$
\cite{BaBarb1pi}. This means that the interference between penguin
and annihilation contributions should be {\it destructive} and the form
factors $V_0^{Bb_1}$ and $V_0^{Ba_1}$ must be of opposite signs.

We also found that the naive relation $\B(B^-\to b_1^0K^-)\sim
{1\over 2}\B(\ov B^0\to b_1^+K^-)$ holds in QCDF. More precisely,
QCDF predicts
 \be
 R_4\equiv {\B(B^-\to b_1^0K^-)\over \B(\ov
B^0\to b_1^+K^-)}=0.51^{+0.01+0.01+0.20}_{-0.01-0.00-0.02} \qquad
 ({\rm expt:}~1.23\pm0.36),
 \en
where the hadronic uncertainty in $R_4$ arises almost entirely
from weak annihilation (contribution from spectator scattering is
negligible). This indicates that the data of $b_1^0K^-$ and
$b_1^+K^-$ can be simultaneously explained only if the weak
annihilation mechanism plays a dominant role in these decays.

\subsubsection{$B\to K_1(1270)(\pi,K), K_1(1400)(\pi,K)$ decays}

It is evident from Table \ref{tab:BRK1} that the central values of
the calculated branching ratios in QCDF for $K_1^-(1270)\pi^+$ and
$K_1^-(1400)\pi^+$ are too small compared to experiment. This is
not surprising as the same phenomenon also occurs in the penguin
dominated $B\to PV$ and $B\to VV$ decays. For example, the default
results for the branching fractions of $B\to K^*\pi$ obtained in
QCDF are in general too small by a factor of $2\sim 3$ compared to
the data \cite{BN}. This suggests the importance of power
corrections due to the non-vanishing $\rho_A$ and $\rho_H$
parameters  or due to possible final-state rescattering effects
from charm intermediate states \cite{CCSfsi}. It has been
demonstrated in \cite{BN} that in the so-called ``S4" scenario
with $\rho_A=1$ and non-vanishing $\phi_A$, the global results for
the $VP$ modes agree better with the data. It has also been shown
in \cite{BenekeVV} that the choice of $\rho_A
e^{i\phi_A}=0.6\,e^{-i40^\circ}$ will allow one to explain the
polarization effects observed in various $B\to VV$ decays. While
large power corrections from weak annihilation seem to be
inevitable for explaining the $K_1\pi$ rates, one issue is that
large weak annihilation may destroy the existing good agreement
for $a_1^+K^-$ and $b_1^+K^-$ modes.

We notice that while  $K_1(1270)\pi$ rates are insensitive to the
mixing angle $\theta_{K_1}$, the branching fractions of
$K_1(1400)\pi$ are smaller for $\theta_{K_1}=-58^\circ$ than that
for $\theta_{K_1}=-37^\circ$ by a factor of $2\sim 3$ due to the
dependence of the $K_1(1400)$ decay constant on $\theta_{K_1}$,
recalling that $f_{K_1(1400)}\sim 112$ (235) MeV for
$\theta_{K_1}=-58^\circ$ ($-37^\circ$) [cf.  Eq. (\ref{eq:fK1})].
The current measurement of $\ov B^0\to K_1^-(1400)\pi^+$ favors a
mixing angle of $-37^\circ$ over $-58^\circ$.

Just as the case of $B\to KK^*$ decays, we find that the branching
ratios of $B\to K_1(1270)K$ and $K_1(1400)K$ modes are of order
$10^{-7}-10^{-8}$ except for the decay $\ov B^0\to \ov
K_1^0(1270)K^0$ which can have a branching ratio of order $2.3\times
10^{-6}$.  The decay modes $K_1^-K^+$ and $K_1^+K^-$ are of
particular interest as they are the only $AP$ modes which receive
contributions solely from weak annihilation.

\begin{table}[t]
\caption{Same as Table \ref{tab:BRa1} except for the decays $B\to
f_1\pi$, $f_1K$, $h_1\pi$ and $h_1K$ with $f_1=f_1(1285),f_1(1420)$
and $h_1=h_1(1170),h_1(1380)$. We use two different sets of mixing
angles, namely, $\theta_{^3P_1}= 27.9^\circ$ and $\theta_{^1P_1}=
25.2^\circ$ (top), corresponding to $\theta_{K_1}=-37^\circ$, and
$\theta_{^3P_1}= 53.2^\circ$, $\theta_{^1P_1}= -18.1^\circ$
(bottom), corresponding to $\theta_{K_1}=-58^\circ$.}
\label{tab:BRf1}
\begin{ruledtabular}
\begin{tabular}{l r  |l  r }
Mode & Theory & Mode & Theory  \\
\hline
 $B^-\to f_1(1285)\pi^-$ & $5.2^{+0.3+1.3+0.7}_{-0.2-1.0-0.2}$ &
 $B^-\to f_1(1420)\pi^-$ & $0.06^{+0.01+0.01+0.00}_{-0.00-0.00-0.00}$ \\
 $\ov B^0\to f_1(1285)\pi^0$ & $0.26^{+0.03+0.14+0.29}_{-0.03-0.07-0.08}$ &
 $\ov B^0\to f_1(1420)\pi^0$ & $0.003^{+0.003+0.002+0.003}_{-0.002-0.001-0.002}$ \\
 $B^-\to f_1(1285)K^-$ & $14.8^{+3.0+7.5+12.4}_{-2.6-3.9-~5.2}$ &
 $B^-\to f_1(1420)K^-$ & $6.0^{+1.7+1.9+9.0}_{-1.5-1.3-3.1}$ \\
 $\ov B^0\to f_1(1285)\ov K^0$ & $14.6^{+2.7+7.5+11.9}_{-2.3-3.9-5.0}$ &
 $\ov B^0\to f_1(1420)\ov K^0$ & $5.5^{+1.6+1.8+8.4}_{-1.3-1.2-2.8}$ \\
 $B^-\to h_1(1170)\pi^-$ & $4.8^{+0.4+0.9+0.8}_{-0.3-0.8-0.7}$ &
 $B^-\to h_1(1380)\pi^-$ & $0.17^{+0.03+0.06+0.04}_{-0.02-0.06-0.03}$ \\
 $\ov B^0\to h_1(1170)\pi^0$ & $0.19^{+0.06+0.05+0.07}_{-0.04-0.03-0.01}$ &
 $\ov B^0\to h_1(1380)\pi^0$ & $0.006^{+0.009+0.005+0.007}_{-0.004-0.004-0.002}$ \\
 $B^-\to h_1(1170)K^-$ & $10.1^{+4.7+2.1+7.3}_{-3.1-1.4-8.1}$ &
 $B^-\to h_1(1380)K^-$ &  $12.7^{+7.1+9.2+211.4}_{-5.1-4.7-~10.8}$ \\
 $\ov B^0\to h_1(1170)\ov K^0$ & $10.1^{+4.2+2.2+7.2}_{-2.8-1.5-8.1}$ &
 $\ov B^0\to h_1(1380)\ov K^0$ & $11.3^{+6.4+8.5+188.5}_{-4.6-4.3-~~9.6}$ \\
 \hline
 $B^-\to f_1(1285)\pi^-$ & $4.6^{+0.2+1.1+0.6}_{-0.2-0.9-0.2}$ &
 $B^-\to f_1(1420)\pi^-$ & $0.59^{+0.06+0.18+0.10}_{-0.05-0.13-0.05}$ \\
 $\ov B^0\to f_1(1285)\pi^0$ & $0.20^{+0.02+0.12+0.24}_{-0.02-0.06-0.06}$ &
 $\ov B^0\to f_1(1420)\pi^0$ & $0.05^{+0.02+0.03+0.04}_{-0.01-0.02-0.02}$ \\
 $B^-\to f_1(1285)K^-$ & $5.2^{+0.9+3.1+~9.1}_{-0.8-1.5-10.0}$ &
 $B^-\to f_1(1420)K^-$ & $13.8^{+4.0+5.6+17.1}_{-3.3-3.2-~6.3}$ \\
 $\ov B^0\to f_1(1285)\ov K^0$ & $5.2^{+0.8+3.2+3.0}_{-0.7-1.5-1.4}$ &
 $\ov B^0\to f_1(1420)\ov K^0$ & $13.1^{+3.7+5.4+16.2}_{-3.0-3.1-~5.9}$ \\
 $B^-\to h_1(1170)\pi^-$ & $1.8^{+0.3+0.3+0.3}_{-0.2-0.3-0.3}$ &
 $B^-\to h_1(1380)\pi^-$ & $2.9^{+0.2+0.6+0.4}_{-0.1-0.6-0.4}$ \\
 $\ov B^0\to h_1(1170)\pi^0$ & $0.16^{+0.08+0.01+0.06}_{-0.05-0.01-0.04}$ &
 $\ov B^0\to h_1(1380)\pi^0$ & $0.04^{+0.00+0.03+0.04}_{-0.00-0.02-0.01}$ \\
 $B^-\to h_1(1170)K^-$ & $11.3^{+5.8+1.9+23.0}_{-3.7-1.1-~8.2}$ &
 $B^-\to h_1(1380)K^-$ & $5.6^{+0.9+2.3+1.4}_{-0.7-1.2-1.9}$ \\
 $\ov B^0\to h_1(1170)\ov K^0$ & $10.9^{+5.3+1.9+21.1}_{-3.4-1.2-~7.7}$ &
 $\ov B^0\to h_1(1380)\ov K^0$ & $5.5^{+0.7+2.4+1.5}_{-0.7-1.2-2.0}$ \\
\end{tabular}
\end{ruledtabular}
\end{table}

\subsubsection{$B\to f_1(\pi,K), h_1(\pi,K)$ decays}
Branching ratios for the decays $B\to f_1\pi$, $f_1K$, $h_1\pi$ and
$h_1K$ with $f_1=f_1(1285),f_1(1420)$ and $h_1=h_1(1170),h_1(1380)$
are shown in Table \ref{tab:BRf1} for two different sets of mixing
angles: (i) $\theta_{^3P_1}= 53.2^\circ$ and $\theta_{^1P_1}=
-18.1^\circ$, corresponding to $\theta_{K_1}=-37^\circ$, and (ii)
$\theta_{^3P_1}= 27.9^\circ$ and $\theta_{^1P_1}= 25.2^\circ$,
corresponding to $\theta_{K_1}=-58^\circ$ (see Sec.
II.A).\footnote{There are predictions for the decay rates of $B\to
f_1\pi$, $f_1K$, $h_1\pi$ and $h_1K$ in \cite{Calderon}. Since the
$f_1$ and $h_1$ states are not specified there, we will not include
them in Table \ref{tab:BRf1} for comparison.}
Their branching ratios are naively expected to be of order
$10^{-6}\sim 10^{-5}$ except for the color-suppressed $f_1\pi^0$
and $h_1\pi^0$ modes which are suppressed relative to the
color-allowed one such as $f_1(1285)\pi^-$ by a factor of
$|a_2/a_1|^2/2\sim {\cal O}(0.03-0.08)$. However, an inspection of
Table \ref{tab:BRf1} shows some exceptions, for example, $\B(B^-\to
f_1(1420)\pi^-)\ll \B(B^-\to f_1(1285)\pi^-)$ for both sets of
$\theta_{^3P_1}$ and $\B(B^-\to h_1(1380)\pi^-)\ll \B(B^-\to
h_1(1170)\pi^-)$ for $\theta_{^1P_1}=25.2^\circ$. These can be
understood as a consequence of interference. The decay amplitudes
for the tree-dominated channels $h_1\pi^-$ are given by
 \be
 A(B^-\to h_1(1380)\pi^-)&\propto& -V_0^{Bh_1}\sin\theta_{^1P_1}
+V_0^{Bh_8}\cos\theta_{^1P_1},  \non \\
 A(B^-\to h_1(1170)\pi^-)&\propto& V_0^{Bh_1}\cos\theta_{^1P_1}
+V_0^{Bh_8}\sin\theta_{^1P_1}.
 \en
Since the form factors $V_0^{Bh_1}$ and $V_0^{Bh_8}$ are of the
same signs (cf. Table \ref{tab:SRFF}), it is clear that the
interference is constructive (destructive) in the $h_1(1170)\pi^-$
mode, but destructive (constructive) in $h_1(1380)\pi^-$ for
$\theta_{^1P_1}=25.2^\circ$ ($-18.1^\circ$). This explains why
$\B(B^-\to h_1(1380)\pi^-)\ll \B(B^-\to h_1(1170)\pi^-)$ for
$\theta_{^1P_1}=25.2^\circ$ and $\B(B^-\to h_1(1380)\pi^-)>
\B(B^-\to h_1(1170)\pi^-)$ for $\theta_{^1P_1}=-18.1^\circ$.
Therefore, a measurement of the ratio $R_5\equiv \B(B^-\to
h_1(1380)\pi^-)/\B(B^-\to h_1(1170)\pi^-)$ will help determine the
mixing angle $\theta_{^1P_1}$. Likewise, information on the angle
$\theta_{^3P_1}$ can be inferred from the ratio $R_6\equiv \B(\ov
B\to f_1(1420)\ov K)/\B(\ov B\to f_1(1285)\ov K)$: $R_6>1$ for
$\theta_{^3P_1}=53.2^\circ$ and  $R_6<1$ for
$\theta_{^3P_1}=27.9^\circ$.

\begin{figure}[t]
\vspace{0cm} \centerline{\epsfig{figure=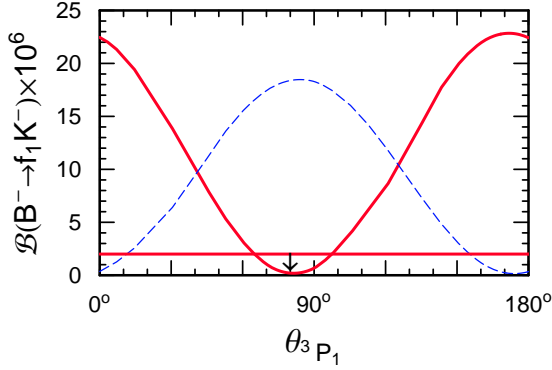,width=8cm}}
    \caption[]{\small Branching ratios of $B^-\to f_1(1285)K^-$ (solid line)
    and $B^-\to f_1(1420)K^-$ (dashed line) versus the
    mixing angle $\theta_{^3P_1}$. The physical mixing angle is
    either $28^\circ$ or $53^\circ$.
    For simplicity, only the central values are shown here.
    The horizontal line is the experimental limit on $B^-\to f_1(1285)K^-$.}
    \label{fig:f1K}
\end{figure}

The preliminary BaBar results are \cite{Burke}
 \be
&& \B(B^-\to f_1(1285)K^-)\B(f_1(1285)\to
\eta\pi\pi)<0.8\times 10^{-6}, \non \\
&& \B(B^-\to f_1(1420)K^-)\B(f_1(1420)\to
\eta\pi\pi)<2.9\times 10^{-6}, \non \\
&& \B(B^-\to f_1(1420)K^-)\B(f_1(1420)\to
K_SK^\pm\pi^\mp)<4.1\times 10^{-6}.
 \en
Since $\B(f_1(1285)\to \eta\pi\pi)=0.52\pm0.16$ \cite{PDG}, the
upper limit on $\B(B^-\to f_1(1285)K^-)$ is inferred to be of order
$2.0\times 10^{-6}$. However, we cannot extract the upper bound for
the $f_1(1420)K^-$ mode due to the lack of information on
$\B(f_1(1420)\to \eta\pi\pi)$ and $\B(f_1(1420)\to
K_SK^\pm\pi^\mp)$. In Fig. \ref{fig:f1K} we plot the branching
ratios of $B^-\to f_1(1285)K^-$ and $B^-\to f_1(1420)K^-$ as a
function of $\theta_{^3P_1}$. We see that the branching fraction of
the former at $\theta_{^3P_1}=53^\circ$ is barely consistent with
the experimental limit when the theoretical errors are taken into
account. Note that the mixing angle dependence of the
$f_1(1420)K^-$ mode is opposite to that of $f_1(1285)K^-$. At this
moment, it is too early to draw any conclusions from the data.
Certainly, we have to await more measurements to test our
predictions.

\subsection{\CP asymmetries}

\begin{table}[t]
\caption{Direct CP asymmetries (in \%) in the decays $B\to
a_1(1260)\, \pi$, $a_1(1260)\, K$, $b_1(1235)\pi$ and
$b_1(1235)K$. See Table \ref{tab:BRa1} for the explanation of
theoretical errors. Experiments results are taken from
\cite{BaBara1pitime,BaBarb1pi,Brown,HFAG}.}
 \label{tab:CPa1}
\begin{ruledtabular}
\begin{tabular}{l r r| l r r}
Mode & Theory & Expt.  & Mode & Theory & Expt.  \\
\hline
 $\overline B^0\to a_1^+ \pi^-$   &
 $-3.6^{+0.1+0.3+20.8}_{-0.1-0.5-20.2}$& $7 \pm 21\pm 15$ \footnotemark[1]
 & $\overline B^0\to a_1^+ K^-$   &  $2.6^{+0.0+0.7+10.1}_{-0.1-0.7-11.0}$
 & $-16\pm12\pm1$  \\
 $\overline B^0\to a_1^-\pi^+$    &
 $-1.9^{+0.0+0.0+14.6}_{-0.0-0.0-14.3}$  & $15\pm15\pm \, 7$ \footnotemark[1]
 &$\overline B^0\to a_1^0 \overline K^0$ & $-7.7^{+0.6+2.1+6.8}_{-0.6-2.2-7.0}$
 \nonumber \\
 $  \overline B^0\to a_1^0 \pi^0$  &
 $60.1^{+4.6+ 6.8+37.6}_{-4.9-8.3-60.7}$&
 &$B^-\to a_1^- \overline K^0$    & $0.8^{+0.0+0.1+0.6}_{-0.0-0.1-0.0}$
 & $12\pm11\pm2$  \\
 $B^-\to a_1^- \pi^0$  & $0.5^{+0.3+0.6+12.0}_{-0.2-0.3-11.0}$ &
 & $B^-\to a_1^0  K^-$             & $8.4^{+0.3+1.4+10.3}_{-0.3-1.6-12.0}$
 \nonumber\\
 $B^-\to a_1^0 \pi^-$ & $-4.3^{+0.3+1.4+14.1}_{-0.3-2.2-14.5}$&  & \nonumber \\
 \hline
 $\overline B^0\to b_1^+ \pi^-$   &
 $-4.0^{+0.2+0.4+26.2}_{-0.0-0.6-25.5}$& $$
 & $\overline B^0\to b_1^+ K^-$   & $5.5^{+0.2+1.2+47.2}_{-0.3-1.2-30.2}$
 & $-7\pm12\pm2$  \\
 $\overline B^0\to b_1^-\pi^+$    &
 $66.1^{+1.2+7.4+30.3}_{-1.4-4.8-96.6}$  & $$
 &$\overline B^0\to b_1^0 \overline K^0$ & $-8.6^{+0.8+3.3+~8.3}_{-0.8-4.2-25.4}$
 \nonumber \\
 $  \overline B^0\to b_1^0 \pi^0$  &
 $53.4^{+6.4+9.0+5.2}_{-6.3-7.3-4.7}$&
 &$B^-\to b_1^- \overline K^0$    & $1.4^{+0.1+0.1+5.6}_{-0.1-0.1-0.1}$
 \nonumber \\
 $B^-\to b_1^- \pi^0$  & $-36.5^{+4.4+18.4+82.2}_{-4.3-17.7-59.6}$ &
 & $B^-\to b_1^0  K^-$ & $18.7^{+1.6+7.8+57.7}_{-1.7-6.1-44.9}$ & $-46\pm20\pm2$
 \nonumber\\
 $B^-\to b_1^0 \pi^-$ & $0.9^{+0.6+2.3+18.0}_{-0.4-2.7-20.5}$& $5\pm16\pm2$  & \nonumber \\
\end{tabular}
\footnotetext[1]{taken from \cite{HFAG}.}
\end{ruledtabular}
\end{table}

\begin{table}[t]
\caption{Direct CP asymmetries (in \%) the decays $B\to K_1(1270)\,
\pi$, $K_1(1270)\, K$, $K_1(1400)\pi$ and $K_1(1400)K$ for two
different mixing angles $\theta_{K_1}=-37^\circ$ (top) and
$-58^\circ$ (bottom).} \label{tab:CPK1}
\begin{ruledtabular}
\begin{tabular}{l r  |l  r }
Mode & Theory & Mode & Theory  \\
\hline
 $\ov B^0\to K_1^-(1270)\pi^+$ & $38.6^{+2.8+10.0+26.0}_{-3.3-13.0-42.9}$ &
 $\ov B^0\to K_1^+(1400)\pi^+$ & $-14.3^{+9.0+1.7+45.3}_{-9.5-1.0-48.8}$ \\
 $\ov B^0\to \ov K_1^0(1270)\pi^0$ & $-32.5^{+1.6+7.4+18.7}_{-1.8-7.6-14.8}$ &
 $\ov B^0\to \ov K_1^0(1400)\pi^0$ & $1.5^{+0.2+0.9+2.3}_{-1.2-0.8-2.2}$ \\
 $B^-\to \ov K_1^0(1270)\pi^-$ & $-0.8^{+0.3+0.1+3.3}_{-0.3-0.2-4.2}$ &
 $B^-\to \ov K_1^0(1400)\pi^-$ & $2.2^{+0.4+0.1+1.5}_{-0.3-0.1-1.2}$ \\
 $\ov B^0\to K_1^-(1270)\pi^0$ & $38.8^{+1.9+7.1+24.5}_{-1.8-9.1-~2.5}$ &
 $\ov B^0\to K_1^-(1400)\pi^0$ & $-12.9^{+7.3+2.2+40.4}_{-6.5-2.7-~9.1}$ \\
 $\ov B^0\to K_1^-(1270)K^+$ & $0^{+0+0+7.3}_{-0-0-7.3}$ &
 $\ov B^0\to K_1^-(1400)K^+$ & $0^{+0+0+13.5}_{-0-0-13.5}$ \\
 $\ov B^0\to K_1^+(1270)K^-$ & $0^{+0+0+40.9}_{-0-0-40.9}$ &
 $\ov B^0\to K_1^+(1400)K^-$ & $0^{+0+0+88.9}_{-0-0-88.9}$ \\
 $\ov B^0\to \ov K_1^0(1270)K^0$ & $-31.0^{+3.5+4.4+10.8}_{-3.1-3.5-11.2}$ &
 $\ov B^0\to \ov K_1^0(1400)K^0$ & $-7.1^{+0.7+14.9+51.1}_{-0.7-~5.1-14.0}$ \\
 $\ov B^0\to K_1^0(1270)\ov K^0$ & $-17.9^{+6.0+0.9+3.9}_{-6.4-0.8-3.0}$ &
 $\ov B^0\to K_1^0(1400)\ov K^0$ & $-65.5^{+4.0+1.4+12.9}_{-4.0-1.4-18.0}$ \\
 $B^-\to K_1^0(1270)K^-$ & $17.2^{+5.8+4.2+65.0}_{-5.4-2.9-~7.6}$ &
 $B^-\to K_1^0(1420)K^-$ & $-45.0^{+7.7+1.5+23.7}_{-7.5-1.0-10.8}$ \\
 $B^-\to K_1^-(1270)K^0$ & $-40.2^{+6.7+2.1+131.3}_{-8.3-5.0-~11.7}$ &
 $B^-\to K_1^-(1420)K^0$ & $-17.3^{+22.1+26.4+101.8}_{-22.1-36.2-~31.2}$ \\
 \hline
 $\ov B^0\to K_1^-(1270)\pi^+$ & $33.6^{+2.6+~8.5+31.2}_{-2.3-10.1-50.7}$ &
 $\ov B^0\to K_1^-(1400)\pi^+$ & $-39.2^{+8.0+1.5+40.7}_{-2.9-2.9-35.4}$ \\
 $\ov B^0\to \ov K_1^0(1270)\pi^0$ & $-29.6^{+1.4+6.8+19.7}_{-1.4-7.9-23.5}$ &
 $\ov B^0\to \ov K_1^0(1400)\pi^0$ & $0.1^{+4.3+1.3+6.0}_{-5.0-1.2-5.6}$ \\
 $B^-\to \ov K_1^0(1270)\pi^-$ & $-0.5^{+0.2+0.0+2.7}_{-0.2-0.2-2.2}$ &
 $B^-\to \ov K_1^0(1400)\pi^-$ & $3.1^{+0.2+0.9+3.0}_{-0.1-0.9-1.7}$ \\
 $\ov B^0\to K_1^-(1270)\pi^0$ & $32.3^{+0.5+5.1+26.5}_{-0.5-6.7-~0.7}$ &
 $\ov B^0\to K_1^-(1400)\pi^0$ & $-42.2^{+7.2+5.6+33.9}_{-8.5-8.7-12.1}$ \\
 $\ov B^0\to K_1^-(1270)K^+$ & $0^{+0+0+27.0}_{-0-0-27.0}$ &
 $\ov B^0\to K_1^-(1400)K^+$ & $0^{+0+0+19.7}_{-0-0-19.7}$ \\
 $\ov B^0\to K_1^+(1270)K^-$ & $0^{+0+0+40.1}_{-0-0-40.1}$ &
 $\ov B^0\to K_1^+(1400)K^-$ & $0^{+0+0+48.1}_{-0-0-48.1}$ \\
 $\ov B^0\to \ov K_1^0(1270)K^0$ & $-18.0^{+0.5+1.2+1.3}_{-0.6-0.7-0.1}$ &
 $\ov B^0\to \ov K_1^0(1400)K^0$ & $-24.3^{+2.2+2.4+10.1}_{-2.0-5.2-56.3}$ \\
 $\ov B^0\to K_1^0(1270)\ov K^0$ & $11.0^{+1.0+2.4+13.0}_{-1.2-2.1-11.0}$ &
 $\ov B^0\to K_1^0(1400)\ov K^0$ & $-63.0^{+4.3+1.7+19.3}_{-3.0-2.4-23.8}$ \\
 $B^-\to K_1^0(1270)K^-$ & $9.4^{+3.1+3.5+41.7}_{-3.3-2.5-~3.3}$ &
 $B^-\to K_1^0(1420)K^-$ & $-59.8^{+2.4+1.2+43.3}_{-2.8-1.4-~3.9}$ \\
 $B^-\to K_1^-(1270)K^0$ & $-39.4^{+6.7+4.3+127.7}_{-6.9-7.8-~~9.7}$ &
 $B^-\to K_1^-(1420)K^0$ & $61.1^{+21.6+23.2+36.3}_{-18.0-41.8-25.5}$ \\
\end{tabular}
\end{ruledtabular}
\end{table}

\begin{table}[t]
\caption{Direct CP asymmetries (in \%) in  the decays $B\to f_1\pi$,
$f_1K$, $h_1\pi$ and $h_1K$ with $f_1=f_1(1285),f_1(1420)$ and
$h_1=h_1(1170),h_1(1380)$. We use two different sets of mixing
angles, namely, $\theta_{^3P_1}= 27.9^\circ$ and $\theta_{^1P_1}=
25.2^\circ$ (top), corresponding to $\theta_{K_1}=-37^\circ$, and
$\theta_{^3P_1}= 53.2^\circ$ and $\theta_{^1P_1}= -18.1^\circ$
(bottom), corresponding to $\theta_{K_1}=-58^\circ$.}
\label{tab:CPf1}
\begin{ruledtabular}
\begin{tabular}{l r  |l  r }
Mode & Theory & Mode & Theory  \\
\hline
 $B^-\to f_1(1285)\pi^-$ & $-7.3^{+0.4+0.5+28.0}_{-0.5-0.6-27.5}$ &
 $B^-\to f_1(1420)\pi^-$ & $-4.1^{+0.9+0.6+44.6}_{-1.1-0.7-44.2}$ \\
 $\ov B^0\to f_1(1285)\pi^0$ & $13.8^{+1.7+3.2+52.6}_{-1.7-3.8-58.7}$ &
 $\ov B^0\to f_1(1420)\pi^0$ & $-34.0^{+10.3+8.0+112.2}_{-~8.1-3.3-~66.4}$ \\
 $B^-\to f_1(1285)K^-$ & $2.5^{+0.3+0.8+6.7}_{-0.2-0.7-8.0}$ &
 $B^-\to f_1(1420)K^-$ & $0.8^{+0.1+0.1+1.6}_{-0.1-0.1-1.7}$ \\
 $\ov B^0\to f_1(1285)\ov K^0$ & $1.9^{+0.1+0.4+2.0}_{-0.1-0.4-2.6}$ &
 $\ov B^0\to f_1(1420)\ov K^0$ & $1.0^{+0.1+0.2+0.9}_{-0.2-0.2-0.8}$ \\
 $B^-\to h_1(1170)\pi^-$ & $-11.1^{+1.1+1.8+3.4}_{-1.1-2.3-3.7}$ &
 $B^-\to h_1(1380)\pi^-$ & $-18.2^{+2.1+3.6+23.3}_{-2.3-5.4-23.3}$ \\
 $\ov B^0\to h_1(1170)\pi^0$ & $31.6^{+7.5+6.5+58.8}_{-6.8-7.5-76.7}$ &
 $\ov B^0\to h_1(1380)\pi^0$ & $-38.7^{+13.9+13.9+124.1}_{-18.7-~9.7-~72.4}$ \\
 $B^-\to h_1(1170)K^-$ & $3.5^{+0.8+0.3+14.1}_{-0.7-0.2-16.5}$ &
 $B^-\to h_1(1380)K^-$ &  $1.3^{+0.2+0.8+12.8}_{-0.5-0.6-18.2}$ \\
 $\ov B^0\to h_1(1170)\ov K^0$ & $1.7^{+0.2+0.2+2.9}_{-0.1-0.1-2.7}$ &
 $\ov B^0\to h_1(1380)\ov K^0$ & $1.8^{+0.3+0.4+0.4}_{-0.3-0.5-3.7}$ \\
 \hline
 $B^-\to f_1(1285)\pi^-$ & $-7.1^{+0.4+0.5+28.6}_{-0.4-0.6-28.0}$ &
 $B^-\to f_1(1420)\pi^-$ & $-3.8^{+0.3+0.4+26.4}_{-0.4-0.4-26.0}$ \\
 $\ov B^0\to f_1(1285)\pi^0$ & $14.7^{+1.7+3.0+57.0}_{-1.9-3.8-64.0}$ &
 $\ov B^0\to f_1(1420)\pi^0$ & $21.0^{+3.2+7.0+40.2}_{-2.7-6.5-45.0}$ \\
 $B^-\to f_1(1285)K^-$ & $2.5^{+0.3+0.9+3.1}_{-0.3-0.7-1.4}$ &
 $B^-\to f_1(1420)K^-$ & $1.3^{+0.3+0.3+3.4}_{-0.2-0.3-3.5}$ \\
 $\ov B^0\to f_1(1285)\ov K^0$ & $2.1^{+0.2+0.6+2.5}_{-0.1-0.5-3.1}$ &
 $\ov B^0\to f_1(1420)\ov K^0$ & $1.1^{+0.2+0.2+1.1}_{-0.1-0.2-1.2}$ \\
 $B^-\to h_1(1170)\pi^-$ & $-8.4^{+0.7+1.3+7.2}_{-0.8-1.7-7.9}$ &
 $B^-\to h_1(1380)\pi^-$ & $-13.8^{+1.4+2.3+1.4}_{-1.3-2.9-1.4}$ \\
 $\ov B^0\to h_1(1170)\pi^0$ & $24.9^{+7.2+2.8+46.5}_{-5.6-3.2-52.1}$ &
 $\ov B^0\to h_1(1380)\pi^0$ & $56.0^{+3.7+11.2+~53.5}_{-4.1-13.6-116.3}$ \\
 $B^-\to h_1(1170)K^-$ & $5.0^{+2.2+0.8+13.7}_{-1.4-0.9-12.4}$ &
 $B^-\to h_1(1380)K^-$ & $-8.1^{+0.4+3.6+4.4}_{-0.2-3.5-4.0}$ \\
 $\ov B^0\to h_1(1170)\ov K^0$ & $0.7^{+0.2+0.2+2.1}_{-0.2-0.2-2.1}$ &
 $\ov B^0\to h_1(1380)\ov K^0$ & $1.5^{+0.5+0.3+1.6}_{-0.6-0.5-1.5}$ \\
\end{tabular}
\end{ruledtabular}
\end{table}

Direct \CP asymmetries for various $B\to AP$ decays are summarized
in Tables \ref{tab:CPa1}-\ref{tab:CPf1}. Due to the large
suppression of $\ov B^0\to b_1^-\pi^+$ relative to $\ov B^0\to
b_1^+\pi^-$, direct CP violation in the later should be close to
the charge asymmetry $\A_{b_1\pi}$ defined below in Eq.
(\ref{eq:chargeA}) which has been measured by BaBar to be
$-0.05\pm0.10\pm0.02$ \cite{BaBarb1pi}. The default results for
direct \CP violation vanishes in the decays $B^0\to K_1^+K^-$ and
$B^0\to K_1^-K^+$ (see Table \ref{tab:CPK1}) as they proceed only through
weak annihilation. The major uncertainty with direct \CP violation
comes from the strong phases which are needed to induce partial
rate \CP asymmetries. For penguin dominated decays, one of the
main sources of strong phases comes from $\phi_A$ defined in Eq.
(\ref{eq:XA}) which is originated from soft gluon interactions. It
is nonperturbative in nature and hence not calculable.

The experimental determination of direct \CP asymmetries for
$a_1^+\pi^-$ and $a_1^-\pi^+$ is more complicated as $B^0\to
a_1^\pm\pi^\mp$ is not a \CP eigenstate. The time-dependent \CP
asymmetries are given by
 \be
 \A(t) &\equiv & {\Gamma(\ov
B^0(t)\to a_1^\pm\pi^\mp)-\Gamma(B^0(t)\to a_1^\pm\pi^\mp)\over
\Gamma(\ov B^0(t)\to a_1^\pm\pi^\mp)+\Gamma(B^0(t)\to
a_1^\pm\pi^\mp)} \non \\
 &=& (S\pm \Delta S)\sin(\Delta
 mt)-(C\pm \Delta C)\cos(\Delta mt),
 \en
where $\Delta m$ is the mass difference of the two neutral $B$
eigenstates, $S$ is referred to as mixing-induced \CP asymmetry
and $C$ is the direct \CP asymmetry, while $\Delta S$ and $\Delta
C$ are {\it CP}-conserving quantities. Defining
\begin{eqnarray}
 A_{+-} & \equiv & A(B^0\to a_1^+\pi^-)~,~~~
A_{-+}  \equiv A(B^0\to a_1^-\pi^+)~,\nonumber\\
\bar{A}_{-+} & \equiv & A(\overline B^0\to a_1^-\pi^+)~,~~~
\bar{A}_{+-} \equiv A(\overline B^0 \to a_1^+\pi^-),
\end{eqnarray}
and
 \be
 \lambda_{+-}={q\over p}\,{\bar A_{+-}\over A_{+-}}, \qquad
 \lambda_{-+}={q\over p}\,{\bar A_{-+}\over A_{-+}},
 \en
where $q/p=e^{-2i\beta}$ for $a_1\pi$ modes, we have
 \be \label{eq:C}
 C+\Delta C={1-|\lambda_{+-}|^2\over 1+|\lambda_{+-}|^2}=
 {|A_{+-}|^2-|\bar A_{+-}|^2\over |A_{+-}|^2+|\bar A_{+-}|^2}, \quad
 C-\Delta C={1-|\lambda_{-+}|^2\over 1+|\lambda_{-+}|^2}={|A_{-+}|^2-|\bar A_{-+}|^2\over |A_{-+}|^2+|\bar A_{-+}|^2},
 \en
and
 \be
 S+\Delta S\equiv {2\,{\rm Im}\lambda_{+-}\over 1+|\lambda_{+-}|^2}={2\,{\rm Im}(e^{-2i\beta}\bar A_{+-}A_{+-}^*)\over
 |A_{+-}|^2+|\bar A_{+-}|^2}, \non \\
  S-\Delta S\equiv {2\,{\rm Im}\lambda_{-+}\over 1+|\lambda_{-+}|^2}={2\,{\rm Im}(e^{-2i\beta}\bar A_{-+}A_{-+}^*)\over
 |A_{-+}|^2+|\bar A_{-+}|^2}.
 \en
Hence we see that $\Delta S$ describes the strong phase difference
between the amplitudes contributing to $B^0\to a_1^\pm\pi^\mp$ and
$\Delta C$ measures the asymmetry between $\Gamma(B^0\to
a_1^+\pi^-)+\Gamma(\ov B^0\to a_1^-\pi^+)$ and $\Gamma(B^0\to
a_1^-\pi^+)+\Gamma(\ov B^0\to a_1^+\pi^-)$.

\begin{table}[t]
\caption{\label{tab:timeCP} Various \CP parameters for the decays
$B^0\to a_1^\pm \pi^\mp$ (top) and $B^0\to b_1^\pm\pi^\mp$
(bottom). The parameters $S$ and $\Delta S$ are computed for
$\beta=22.1^\circ$ and $\gamma =68.0^\circ$. Experimental results
are taken from \cite{BaBara1pitime,BaBarb1pi}.} \vspace{0.0cm}
\begin{ruledtabular}
\begin{tabular}{|c|r r|}
\multicolumn{1}{|c|}{Parameter} & Theory  & Experiment \\
\hline
 $\A_{a_1\pi}$ &
 $\phantom{-}0.003^{\,+0.001\,+0.002\,+0.043}_{\,-0.002\,-0.003\,-0.045}$
 & $-0.07\pm0.07\pm0.02$\\
 $C$ & $\phantom{-}0.02^{\,+0.00\,+0.00\,+0.14}_{\,-0.00\,-0.00\,-0.14}$ &
   $-0.10\pm0.15\pm0.09$ \\
$S$ &
 $ -0.37^{\,+0.01\,+0.05\,+0.09}_{\,-0.01\,-0.08\,-0.16}$ &
   $0.37\pm0.21\pm0.07$ \\
$\Delta C$ &
$\phantom{-}0.44^{\,+0.03\,+0.03\,+0.03}_{\,-0.04\,-0.05\,-0.04}$ &
   $0.26\pm0.15\pm0.07$ \\
 $\Delta S$ & $0.01^{\,+0.00\,+0.00\,+0.02}_{\,-0.00\, -0.00\,-0.02}$ &
   $-0.14\pm0.21\pm0.06$ \\
 & & \\
 $\alpha_{\rm eff}^+$  & $(97.2^{\, +0.3\, +1.0\, +4.7}_{\,-0.3\, -0.6\, -2.5})^\circ$
 &  \\
 $\alpha_{\rm eff}^-$  & $(107.0^{\, +0.5\, +3.6\, +6.6}_{\,-0.5\, -2.3\, -3.7})^\circ$
 &  \\
 $\alpha_{\rm eff}$  & $(102.0^{\, +0.4\, +2.3\, +5.7}_{\,-0.4\, -1.5\, -3.1})^\circ$
 &  $(78.6\pm 7.3)^\circ$ \\
\hline
 $\A_{b_1\pi}$ &
 $-0.06^{\,+0.01\,+0.01\,+0.23}_{\,-0.01\,-0.01\,-0.23}$
 & $-0.05\pm0.10\pm0.02$\\
 $C$ & $-0.03^{\,+0.01\,+0.01\,+0.06}_{\,-0.02\,-0.02\,-0.01}$ &
   $0.22\pm0.23\pm0.05$ \footnotemark[1] \\
$S$ &
 $ 0.05^{\,+0.03\,+0.02\,+0.15}_{\,-0.03\,-0.02\,-0.26}$ &
   $$ \\
$\Delta C$ & $-0.96^{\,+0.03\,+0.02\,+0.08}_{\,-0.03\,-0.03\,-0.01}$
& $-1.04\pm0.23\pm0.08$ \\
 $\Delta S$ & $0.12^{\,+0.04\,+0.04\,+0.08}_{\,-0.03\, -0.04\,-0.09}$ &
   $$ \\
 & & \\
 $\alpha_{\rm eff}^+$  & $(107.6^{\, +0.7\, +3.5\, +155.4}_{\,-0.2\, -4.9\, -~17.8})^\circ$
 &  \\
 $\alpha_{\rm eff}^-$  & $(101.3^{\, +0.4\, +2.1\, +4.9}_{\,-0.4\, -1.4\, -8.6})^\circ$
 &  \\
 $\alpha_{\rm eff}$  & $(104.4^{\, +0.6\, +2.6\, +80.4}_{\,-0.3\, -2.1\, -~1.6})^\circ$
 &  $$ \\
\end{tabular}
\footnotetext[1]{Our definition of $C$ in Eq. (\ref{eq:C}) has an
opposite sign to that defined in \cite{BaBarb1pi} for $B\to
b_1\pi$ decays.}
\end{ruledtabular}
\end{table}

Next consider the time- and flavor-integrated charge asymmetry
 \be \label{eq:chargeA}
 \A_{a_1\pi}\equiv {|A_{+-}|^2+|\bar A_{+-}|^2-|A_{-+}|^2-|\bar
 A_{-+}|^2\over |A_{+-}|^2+|\bar A_{+-}|^2+|A_{-+}|^2+|\bar
 A_{-+}|^2},
 \en
Then, following \cite{CKMfitter} one can transform the
experimentally motivated \CP parameters $\A_{a_1\pi}$ and
$C_{a_1\pi}$ into the physically motivated choices
 \be
 A_{a_1^+\pi^-} &\equiv& {|\kappa^{-+}|^2-1\over |\kappa^{-+}|^2+1},
 \qquad  A_{a_1^-\pi^+} \equiv {|\kappa^{+-}|^2-1\over |\kappa^{+-}|^2+1},
 \en
with
 \be
 \kappa^{+-}={q\over p}\,{\bar A_{-+}\over A_{+-}}, \qquad
  \kappa^{-+}={q\over p}\,{\bar A_{+-}\over A_{-+}}.
  \en
Hence,
 \be
 A_{a_1^+\pi^-} &=& {\Gamma(\ov B^0\to a_1^+\pi^-)-\Gamma(B^0\to
 a_1^-\pi^+)\over \Gamma(\ov B^0\to a_1^+\pi^-)+\Gamma(B^0\to
 a_1^-\pi^+)}={\A_{a_1\pi}-C_{a_1\pi}-\A_{a_1\pi}\Delta C_{a_1\pi}\over 1-\Delta
 C_{a_1\pi}-\A_{a_1\pi} C_{a_1\pi}},   \non \\
 A_{a_1^-\pi^+} &=& {\Gamma(\ov B^0\to a_1^-\pi^+)-\Gamma(B^0\to
 a_1^+\pi^-)\over \Gamma(\ov B^0\to a_1^-\pi^+)+\Gamma(B^0\to
 a_1^+\pi^-)}=-{\A_{a_1\pi}+C_{a_1\pi}+\A_{a_1\pi}\Delta C_{a_1\pi}\over 1+\Delta
 C_{a_1\pi}+\A_{a_1\pi} C_{a_1\pi}}.
 \en
Note that the quantities $A_{a_1^\pm\pi^\mp}$ here correspond to
$A_{a_1^\mp\pi^\pm}$ defined in \cite{CKMfitter}. Therefore,
direct \CP asymmetries $A_{a_1^+\pi^-}$ and $A_{a_1^-\pi^+}$ are
determined from the above two equations.

Defining the effective phases
 \be
 \alpha_{\rm eff}^{+} &\equiv& {1\over 2}\,{\rm arg}\kappa^{+-}={1\over
 2}\,{\rm arg}\left(e^{-2i\beta}\bar A_{-+}A_{+-}^*\right), \non
 \\   \alpha_{\rm eff}^{-} &\equiv& {1\over 2}\,{\rm arg}\kappa^{-+}={1\over
 2}\,{\rm arg}\left(e^{-2i\beta}\bar A_{+-}A_{-+}^*\right),
 \en
which reduce to the unitarity angle $\alpha$ in the absence of
penguin contributions, we have
 \begin{eqnarray}\label{eq:alpha-eff}
 \alpha_{\rm eff}
 &\equiv& \frac{1}{2}\Big(\alpha^+_{\rm eff} + \alpha^-_{\rm eff}
 \Big) \nonumber\\
 &=& \frac{1}{4} \Bigg[ \arcsin \Bigg(\frac{S + \Delta S}
    {\sqrt{1- (C + \Delta C)^2}}\Bigg)
 + \arcsin\Bigg(\frac{S - \Delta S}{\sqrt{1- (C - \Delta C)^2}}\Bigg) \Bigg]~.
\end{eqnarray}
This is a measurable quantity which is equal to the weak phase
$\alpha$ in the limit of vanishing penguin amplitudes.

Parameters of the time-dependent decay rate asymmetries of $B^0\to
a_1^\pm \pi^\mp$  are shown in Table \ref{tab:timeCP}. It
appears that the calculated mixing-induced parameter $S$ is
negative and the effective unitarity angle $\alpha_{\rm eff}$
deviates from experiment by around $2\sigma$. As pointed out by
one of us (K.C.Y.), this discrepancy may be resolved by having a
larger $\gamma\gsim 80^\circ$ (see Fig. 1 of \cite{Yanga1}).
Further precise measurements are needed to clarify the
discrepancy. For $B^0\to b_1^\pm\pi^\mp$ decays, the predicted
$\Delta C$ agrees with experiment. The fact that this quantity is
very close to $-1$ indicates that the $\ov B^0\to b_1^-\pi^+$ mode
is highly suppressed relative to the $\ov B^0\to b_1^+\pi^-$ one,
recalling that $\Delta C$ here measures the asymmetry between
$\Gamma(B^0\to b_1^+\pi^-)+\Gamma(\ov B^0\to b_1^-\pi^+)$ and
$\Gamma(B^0\to b_1^-\pi^+)+\Gamma(\ov B^0\to b_1^+\pi^-)$.


\subsection{Comparison with other works}
There are several papers  studying charmless $B\to AP$ decays:
Laporta, Nardulli, and Pham (LNP) \cite{Nardulli07} (see also
Nardulli and Pham \cite{Nardulli05}), and Calder\'on, Mu\~noz and
Vera (CMV) \cite{Calderon}. Their predictions are shown in Tables
\ref{tab:BRa1} and \ref{tab:BRK1}. Both are based on naive
factorization. While form factors are obtained by CMV using the
ISGW2 model, LNP use ratios of branching ratios to deduce ratios
of form factors. Hence, the relevant form factors are determined
by factorization and experimental data. Specifically, LNP found
 \be \label{eq:FFLNP}
 {V_0^{Ba_1}(0)\over A_0^{B\rho}(0)}\approx {V_0^{BK_{1A}}(0)\over
 A_0^{BK^*}(0)}
 &=& h\sin\theta_{K_1}+k\cos\theta_{K_1}, \non \\
 {V_0^{Bb_1}(0)\over A_0^{B\rho}(0)}\approx {V_0^{BK_{1B}}(0)\over
 A_0^{BK^*}(0)}
 &=& h\cos\theta_{K_1}-k\sin\theta_{K_1},
 \en
where the kinematic factors $h,k$ are defined in \cite{Nardulli07}.
Therefore, LNP used two different sets of form factors corresponding
to different mixing angle values $\theta_{K_1}=32^\circ$ and
$58^\circ$. \footnote{Since  the $B\to K_{1A}$ and $B\to K_{1B}$
form factors obtained in the ISGW2 model are opposite in sign, the
preferred mixing angle $\theta_{K_1}$ should be negative, as
discussed in Sec. II.A. }
Since LNP only considered factorizable contributions to $B\to AP$
decays, it turns out that in $B\to K_1\pi$ decays, only the
$K_1^-(1400)\pi^0$ and $\ov K_1^0(1400)\pi^0$ modes depend on the
mixing angle $\theta_{K_1}$. The other $K_1\pi$ rates obtained by
LNP (see Table \ref{tab:BRK1}) are mixing angle independent.

The predicted rates for $a_1\pi$, $b_1\pi$ and $b_1K$ modes by CMV
are generally too large compared to the data, presumably due to too
big form factors for $B\to a_1 (b_1)$ transition predicted by the
ISGW2 model. The relation $\B(\ov B^0\to a_1^+\pi^-)>\B(\ov B^0\to
a_1^-\pi^+)$ is in conflict with experiment. A noticeable result
found by CMV is that $\B(B^-\to K_1^-(1400)K^0)$ and $\B(\ov B^0\to
\ov K_1^0(1400)K^0)$ are of order $10^{-6}$, while in QCDF they are
highly suppressed, of order $10^{-7}-10^{-8}$.

It is clear from Table \ref{tab:BRa1} that in the  LNP model, the
form factors (\ref{eq:FFLNP}) derived using
$\theta_{K_1}=32^\circ$ give a  better agreement for $a_1\pi$
modes, whereas $\theta_{K_1}=58^\circ$ is preferred by $b_1\pi$
and $b_1K$ data. This indicates that the data of $a_1(\pi,K)$ and
$b_1(\pi,K)$ cannot be simultaneously accounted for by a single
mixing angle $\theta_{K_1}$ in this model.

Branching ratios of $B\to f_1P$ and $B\to h_1P$  are found to be
of order $10^{-5}$ for $P=\pi^\pm,\eta,\eta',K$ and ${\cal
O}(10^{-7})$ for $P=\pi^0$ by CMV. In general, the CMV's
predictions are larger than ours by one order of magnitude.

\section{Conclusions}

In this work we have studied the two-body hadronic decays of $B$
mesons into pseudoscalar and axial vector mesons within the
framework of QCD factorization. The light-cone distribution
amplitudes for $^3P_1$ and $^1P_1$ axial-vector mesons have been
evaluated using the QCD sum rule method. Owing to the $G$-parity,
the chiral-even two-parton light-cone distribution amplitudes of
the $^3P_1$ ($^1P_1$) mesons are symmetric (antisymmetric) under
the exchange of quark and anti-quark momentum fractions in the
SU(3) limit. For chiral-odd light-cone distribution amplitudes, it
is other way around. Our main conclusions are as follows:

\begin{itemize}

\item Using the Gell-Mann-Okubo mass formula and the $K_1(1270)$
and $K_1(1400)$ mixing angle $\theta_{K_1}=-37^\circ(-58^\circ)$,
the mixing angles for $^3P_1$ and $^1P_1$ states are found to be
$\theta_{^3P_1}\sim 28^\circ(53^\circ)$ for the $f_1(1420)$ and
$f_1(1285)$ and $\theta_{^1P_1}\sim 25^\circ(-18^\circ)$ for
$h_1(1170)$ and the $h_1(1380)$, respectively.

\item The predicted rates for $a_1^\pm(1260)\pi^\mp$,
$b_1^\pm(1235)\pi^\mp$, $b_1^0(1235)\pi^-$, $a_1^+K^-$ and
$b_1^+K^-$ modes are in good agreement with the data. However, the
expected ratios $\B(B^-\to a_1^0\pi^-)/\B(\ov B^0\to
a_1^+\pi^-)\lsim 1$, $\B(B^-\to a_1^-\pi^0)/\B(\ov B^0\to
a_1^-\pi^+)\sim {1\over 2}$ and $\B(B^-\to b_1^0K^-)/\B(\ov B^0\to
b_1^+K^-)\sim {1\over 2}$ are not borne out by experiment. This
should be clarified by the improved measurements of these decays
in the future.

\item One of the salient features of the $^1P_1$ axial vector meson
is that its axial-vector decay constant is small, vanishing in the
SU(3) limit. This feature is confirmed by the observation that
$\Gamma(\ov B^0\to b_1^-\pi^+)\ll \Gamma(\ov B^0\to b_1^+\pi^-)$.
By contrast, it is expected that $\Gamma(\ov B^0\to a_1^-\pi^+)\gg
\Gamma(\ov B^0\to a_1^+\pi^-)$ due to the fact that $f_{a_1}\gg
f_\pi$.

\item While $B\to a_1\pi$ decays have similar rates as that of
$B\to \rho\pi$, the penguin-dominated decays $B\to a_1K$ resemble
much more to the $\pi K$ modes than $\rho K$ ones. However, the
naively expected ratio $\B(B^-\to a_1^-\ov K^0)/\B(\ov B^0\to
a_1^+K^-)\approx \B(B^-\to \pi^-\ov K^0)/\B(\ov B^0\to
\pi^+K^-)\approx 1.2$ is not consistent with the current
experimental value of $2.14\pm0.63$\,.

\item Since the $\ov B\to b_1K$ decays receive sizable
annihilation contributions, their rates are sensitive to the
interference between penguin and annihilation terms. The
measurement of $\B(\ov B^0\to b_1^+K^-)$ implies a destructive
interference which in turn indicates that the form factors for
$B\to b_1$ and $B\to a_1$ transitions must be of opposite signs.

\item The central values of the branching ratios for the
penguin-dominated modes $K_1^-(1270)\pi^+$ and $K_1^-(1400)\pi^+$
predicted by QCD factorization are too small compared to experiment.
Just as the case of $B\to K^*\pi$ decays, sizable power corrections
such as weak annihilation are needed to account for the observed
$K_1\pi$ rates. The current measurement of $\ov B^0\to
K_1^-(1400)\pi^+$ seems to favor a $K_{1A}-K_{1B}$ mixing angle of
$-37^\circ$ over $-58^\circ$.

\item The decays $B\to K_1\ov K$ with $K_1=K_1(1270)$ and
$K_1(1400)$ are in general quite suppressed, of order $10^{-7}\sim
10^{-8}$, except for $\ov B^0\to \ov K_1^0(1270)K^0$ which can have
a branching ratio of order $2.3\times 10^{-6}$. The decay modes
$K_1^-K^+$ and $K_1^+K^-$ are of particular interest as they are the
only $AP$ modes that proceed only through weak annihilation.

\item Time-dependent \CP asymmetries in the decays $B^0\to
a_1^\pm\pi^\mp$ and $b_1^\pm\pi^\mp$ are studied. For the former,
the mixing-induced parameter $S$ is found to be negative and the
effective unitarity angle $\alpha_{\rm eff}$ deviates from
experiment by around $2\sigma$. The discrepancy discrepancy
between theory and experiment may be resolved by having a larger
$\gamma\gsim 80^\circ$. Further precise measurements are needed to
clarify the discrepancy.

\item Branching ratios for the decays $B\to f_1\pi$, $f_1K$,
$h_1\pi$ and $h_1K$ with $f_1=f_1(1285),f_1(1420)$ and
$h_1=h_1(1170),h_1(1380)$  are generally of order $10^{-6}$ except
for the color-suppressed $f_1\pi^0$ and $h_1\pi^0$ modes which are
suppressed by one to two orders of magnitude. Measurements of the
ratios $\B(B^-\to h_1(1380)\pi^-)/\B(B^-\to h_1(1170)\pi^-)$ and
$\B(\ov B\to f_1(1420)\ov K)/\B(\ov B\to f_1(1285)\ov K)$ will help
determine the mixing angles $\theta_{^1P_1}$ and $\theta_{^3P_1}$,
respectively.

\end{itemize}

\vskip 2.5cm \acknowledgments

We are grateful to Jim Smith for reading the manuscript and for
valuable comments. This research was supported in part by the
National Science Council of R.O.C. under Grant Nos.
NSC96-2112-M-001-003 and NSC96-2112-M-033-004-MY3.

\newpage
\appendix
\section{Decay amplitudes}

For simplicity, here we do not explicitly show the arguments, $M_1$
and $M_2$, of  $\alpha_i^p$ and $\beta_i^p$ coefficients. The order
of the arguments of the $\alpha_i^p (M_1 M_2)$ and $\beta_i^p(M_1
M_2)$ is consistent with the order of the arguments of the
$X^{(\overline B M_1, M_2)}$, where
 \be \label{eq:beta}
 \beta_i^p (M_1 M_2) =\frac{-i f_B f_{M_1}
f_{M_2}}{X^{(\overline B M_1,M_2)}}b_i^p.
 \en
Within the framework of QCD factorization \cite{BBNS}, the $B\to
AP$ decay amplitudes read
 \begin{eqnarray} \label{eq:a1pi}
 \sqrt2\,{\cal A}_{B^-\to a_1^0 \pi^-}
   &=&  \frac{G_F}{\sqrt{2}}\sum_{p=u,c}\lambda_p^{(d)}
   \Bigg\{ \left[ \delta_{pu}\,(\alpha_2 - \beta_2)
    - \alpha_4^p + \frac{3}{2}\alpha_{3,{\rm EW}}^p
    + \frac{1}{2}\alpha_{4,{\rm EW}}^p  - \beta_3^p - \beta_{3,{\rm EW}}^p
    \right] X^{(\overline B \pi, a_1)} \nonumber\\
   &+&  \left[ \delta_{pu}\,(\alpha_1 + \beta_2)
    + \alpha_4^p + \alpha_{4,{\rm EW}}^p + \beta_3^p
    + \beta_{3,{\rm EW}}^p \right] X^{(\overline B a_1, \pi)}\Bigg\}  , \nonumber\\
 \sqrt2\,{\cal A}_{B^-\to a_1^- \pi^0}
   &=&  \frac{G_F}{\sqrt{2}}\sum_{p=u,c}\lambda_p^{(d)}
   \Bigg\{ \left[ \delta_{pu}\,(\alpha_2 - \beta_2)
    - \alpha_4^p + \frac{3}{2}\alpha_{3,{\rm EW}}^p
    + \frac{1}{2}\alpha_{4,{\rm EW}}^p  - \beta_3^p - \beta_{3,{\rm EW}}^p
    \right] X^{(\overline B a_1, \pi)} \nonumber\\
   &+&  \left[ \delta_{pu}\,(\alpha_1 + \beta_2)
    + \alpha_4^p + \alpha_{4,{\rm EW}}^p + \beta_3^p
    + \beta_{3,{\rm EW}}^p \right] X^{(\overline B \pi, a_1)}\Bigg\}  , \nonumber\\
 {\cal A}_{\bar B^0\to a_1^- \pi^+}
   &=& \frac{G_F}{\sqrt{2}}\sum_{p=u,c}\lambda_p^{(d)}
   \Bigg\{ \left[ \delta_{pu}\,\alpha_1 + \alpha_4^p
    + \alpha_{4,{\rm EW}}^p + \beta_3^p + \beta_4^p
    - \frac{1}{2}\beta_{3,{\rm EW}}^p - \frac{1}{2}\beta_{4,{\rm EW}}^p \right]
    X^{(\overline B \pi, a_1)}
    \nonumber\\
    &+& \left[ \delta_{pu}\,\beta_1 + \beta_4^p
    + \beta_{4,{\rm EW}}^p \right] X^{(\overline B a_1, \pi)}\Bigg\}   , \nonumber\\
 {\cal A}_{\bar B^0\to a_1^+ \pi^-}
   &=& \frac{G_F}{\sqrt{2}}\sum_{p=u,c}\lambda_p^{(d)}
   \Bigg\{ \left[ \delta_{pu}\,\alpha_1 + \alpha_4^p
    + \alpha_{4,{\rm EW}}^p + \beta_3^p + \beta_4^p
    - \frac{1}{2}\beta_{3,{\rm EW}}^p - \frac{1}{2}\beta_{4,{\rm EW}}^p \right]
    X^{(\overline B a_1, \pi)}
    \nonumber\\
   &+& \left[ \delta_{pu}\,\beta_1 + \beta_4^p
    + \beta_{4,{\rm EW}}^p \right]  X^{(\overline B \pi, a_1)}\Bigg\}  , \nonumber\\
 -2\,{\cal A}_{\bar B^0\to \pi^0 a_1^0}
   &=& \frac{G_F}{\sqrt{2}}\sum_{p=u,c}\lambda_p^{(d)}
   \Bigg\{ \Big[ \delta_{pu}\,(\alpha_2 - \beta_1)
    - \alpha_4^p + \frac{3}{2}\alpha_{3,{\rm EW}}^p
    + \frac{1}{2}\alpha_{4,{\rm EW}}^p - \beta_3^p - 2\beta_4^p
    \nonumber\\[-0.1cm]
    &&\hspace*{1cm}
    +\, \frac{1}{2}\beta_{3,{\rm EW}}^p - \frac{1}{2}\beta_{4,{\rm EW}}^p \Big]
    X^{(\overline B \pi, a_1)}
    \nonumber\\
   &+& \Big[ \delta_{pu}\,(\alpha_2 - \beta_1)
    - \alpha_4^p + \frac{3}{2}\alpha_{3,{\rm EW}}^p
    + \frac{1}{2}\alpha_{4,{\rm EW}}^p - \beta_3^p - 2\beta_4^p
    \nonumber\\[-0.1cm]
    &&\hspace*{1cm}
    + \,\frac{1}{2}\beta_{3,{\rm EW}}^p - \frac{1}{2}\beta_{4,{\rm EW}}^p \Big]
     X^{(\overline B a_1, \pi)} \Bigg\}\,,
\end{eqnarray}
for $\ov B\to a_1\pi$,
\begin{eqnarray} \label{eq:a1K}
 {\cal A}_{B^-\to a_1^-\bar K^0}
   &=&  \frac{G_F}{\sqrt{2}}\sum_{p=u,c}\lambda_p^{(s)}
    \left[ \delta_{pu}\,\beta_2
    + \alpha_4^p - \frac{1}{2}\alpha_{4,{\rm EW}}^p + \beta_3^p
    + \beta_{3,{\rm EW}}^p \right]   X^{(\overline B  a_1,\bar K)}, \nonumber\\
   \sqrt2\,{\cal A}_{B^-\to a_1^0 K^-}
   &=&  \frac{G_F}{\sqrt{2}}\sum_{p=u,c}\lambda_p^{(s)}
   \Bigg\{ \left[ \delta_{pu}\,(\alpha_1+\beta_2)
    + \alpha_4^p + \alpha_{4,{\rm EW}}^p + \beta_3^p
    + \beta_{3,{\rm EW}}^p \right] X^{(\overline B  a_1,\bar K)} \nonumber\\
   &+&  \left[ \delta_{pu}\,\alpha_2
    + \frac{3}{2}\alpha_{3,{\rm EW}}^p \right] X^{(\overline B \bar K, a_1)}\Bigg\}  ,
    \nonumber\\
 {\cal A}_{\bar B^0\to a_1^+ K^-}
   &=&  \frac{G_F}{\sqrt{2}}\sum_{p=u,c}\lambda_p^{(s)}
    \left[ \delta_{pu}\,\alpha_1
    + \alpha_4^p + \alpha_{4,{\rm EW}}^p + \beta_3^p
    - \frac{1}{2}\beta_{3,{\rm EW}}^p \right] X^{(\overline B  a_1,\bar K)} ,
    \nonumber\\
 \sqrt2\,{\cal A}_{\bar B^0\to a_1^0\bar K^0}
   &=&  \frac{G_F}{\sqrt{2}}\sum_{p=u,c}\lambda_p^{(s)}
   \Bigg\{ \left[ -\alpha_4^p + \frac{1}{2}\alpha_{4,{\rm EW}}^p
    - \beta_3^p + \frac{1}{2}\beta_{3,{\rm EW}}^p \right]X^{(\overline B  a_1, \bar K)}
  \nonumber\\
   &+&  \left[ \delta_{pu}\,\alpha_2
    + \frac{3}{2}\alpha_{3,{\rm EW}}^p \right]X^{(\overline B \bar K, a_1)} \Bigg\} ,
\end{eqnarray}
and $\ov B\to a_1\ov K$,
\begin{eqnarray}
 \sqrt{2}{\cal A}_{B^-\to f_1^0\pi^-}
   &=&  \frac{G_F}{\sqrt{2}}\sum_{p=u,c}\lambda_p^{(d)} \Bigg\{
    \Big[ \delta_{pu}(\alpha_2
    + \beta_2)+ 2\alpha_3^p +\alpha_4^p + \frac{1}{2}\alpha_{3,{\rm
    EW}}^p  \non \\ &-&  \frac{1}{2}\alpha_{4,{\rm EW}}^p
    + \beta_3^p + \beta_{3,{\rm EW}}^p \Big]   X^{(\overline B \pi,f_1^q)}
    + \sqrt{2}\Big[\alpha_3^p-{1\over 2}\alpha^p_{3,{\rm EW}}\Big]X^{(\overline B \pi,f_1^s)}, \nonumber\\
    &+& \Big[  \delta_{pu}(\alpha_1
    + \beta_2) +\alpha_4^p + \alpha_{4,{\rm EW}}^p
    + \beta_3^p + \beta_{3,{\rm EW}}^p \Big]   X^{(\overline Bf_1^q,
    \pi)}\Bigg\}, \non \\
  -2{\cal A}_{\ov B^0\to f_1^0\pi^0}
   &=&  \frac{G_F}{\sqrt{2}}\sum_{p=u,c}\lambda_p^{(d)} \Bigg\{
    \Big[ \delta_{pu}(\alpha_2
    - \beta_1)+ 2\alpha_3^p +\alpha_4^p + \frac{1}{2}\alpha_{3,{\rm
    EW}}^p  \non \\ &-&  \frac{1}{2}\alpha_{4,{\rm EW}}^p
    + \beta_3^p -{1\over 2} \beta_{3,{\rm EW}}^p -{3\over 2} \beta_{4,{\rm EW}}^p \Big]   X^{(\overline B \pi,f_1^q)}
    + \sqrt{2}\Big[\alpha_3^p-{1\over 2}\alpha^p_{3,{\rm EW}}\Big]X^{(\overline B \pi,f_1^s)}, \nonumber\\
    &+& \Big[  \delta_{pu}(-\alpha_2
    - \beta_1) +\alpha_4^p -{3\over 2}\alpha_{3,{\rm EW}}^p- {1\over 2}\alpha_{4,{\rm EW}}^p
    + \beta_3^p -{1\over 2} \beta_{3,{\rm EW}}^p -{3\over 2} \beta_{4,{\rm EW}}^p\Big]   X^{(\overline Bf_1^q,
    \pi)}\Bigg\},\non \\
  \sqrt{2}{\cal A}_{B^-\to f_1^0 K^-}
   &=&  \frac{G_F}{\sqrt{2}}\sum_{p=u,c}\lambda_p^{(s)} \Bigg\{
    \Big[ \delta_{pu}\alpha_2+ 2\alpha_3^p + \frac{1}{2}\alpha_{3,{\rm
    EW}}^p \Big]   X^{(\overline B K,f_1^q)}  \non \\
    &+& \sqrt{2}\Big[ \delta_{pu}\beta_2 +\alpha_3^p +\alpha_4^p -{1\over 2}\alpha^p_{3,{\rm EW}}
   -{1\over 2}\alpha^p_{4,{\rm EW}} +\beta_3^p +\beta^p_{3,{\rm
   EW}}\Big] X^{(\overline B K,f_1^s)}, \nonumber\\
    &+& \Big[  \delta_{pu}(\alpha_1
    + \beta_2) +\alpha_4^p + \alpha_{4,{\rm EW}}^p
    + \beta_3^p + \beta_{3,{\rm EW}}^p \Big]   X^{(\overline Bf_1^q,
    K)}\Bigg\},\non \\
  \sqrt{2}{\cal A}_{\ov B^0\to f_1^0 \ov K^0}
   &=&  \frac{G_F}{\sqrt{2}}\sum_{p=u,c}\lambda_p^{(s)} \Bigg\{
    \Big[ \delta_{pu}\alpha_2+ 2\alpha_3^p + \frac{1}{2}\alpha_{3,{\rm
    EW}}^p \Big]   X^{(\overline B K,f_1^q)}  \non \\
    &+& \sqrt{2}\Big[ \alpha_3^p +\alpha_4^p -{1\over 2}\alpha^p_{3,{\rm EW}}
   -{1\over 2}\alpha^p_{4,{\rm EW}} +\beta_3^p -{1\over 2}\beta^p_{3,{\rm
   EW}}\Big] X^{(\overline B K,f_1^s)}, \nonumber\\
    &+& \Big[   \alpha_4^p -{1\over 2} \alpha_{4,{\rm EW}}^p
    + \beta_3^p -{1\over 2} \beta_{3,{\rm EW}}^p \Big]   X^{(\overline Bf_1^q,
    K)}\Bigg\},
    \en
for $\ov B\to f_1^0\pi$ and $\ov B\to f_1^0\ov K$,
 \be
  {\cal A}_{B^-\to \ov K^0_1\pi^-}
   &=&  \frac{G_F}{\sqrt{2}}\sum_{p=u,c}\lambda_p^{(s)}
    \Big[ \delta_{pu}\beta_2+ \alpha_4^p - \frac{1}{2}\alpha_{4,{\rm
    EW}}^p +\beta_3^p +\beta^p_{3,{\rm EW}} \Big]   X^{(\overline B\pi, \ov K_1)},  \non \\
  \sqrt{2}{\cal A}_{B^-\to K^-_1\pi^0}
   &=&  \frac{G_F}{\sqrt{2}}\sum_{p=u,c}\lambda_p^{(s)} \Bigg\{
    \Big[ \delta_{pu}(\alpha_1+\beta_2)+ \alpha_4^p +\alpha_{4,{\rm
    EW}}^p +\beta_3^p +\beta^p_{3,{\rm EW}} \Big]   X^{(\overline B \pi,\ov K_1)}  \non \\
   &+& \Big[\delta_{pu}\alpha_2+{3\over 2}\alpha^p_{3,{\rm EW}}\Big]  X^{(\overline B\ov K_1,\pi)}\Bigg\}, \non
   \\
   {\cal A}_{\ov B^0\to K^-_1\pi^+}
   &=&  \frac{G_F}{\sqrt{2}}\sum_{p=u,c}\lambda_p^{(s)}
    \Big[ \delta_{pu}\alpha_1+ \alpha_4^p +\alpha_{4,{\rm
    EW}}^p +\beta_3^p -{1\over 2}\beta^p_{3,{\rm EW}} \Big]   X^{(\overline B \pi,\ov K_1)},  \non \\
   \sqrt{2}{\cal A}_{\ov B^0\to \ov K^0_1\pi^0}
   &=&  \frac{G_F}{\sqrt{2}}\sum_{p=u,c}\lambda_p^{(s)} \Bigg\{
    \Big[ -\alpha_4^p +{1\over 2}\alpha_{4,{\rm
    EW}}^p -\beta_3^p +{1\over 2}\beta^p_{3,{\rm EW}} \Big]   X^{(\overline B \pi,\ov K_1)}  \non \\
   &+& \Big[\delta_{pu}\alpha_2+{3\over 2}\alpha^p_{3,{\rm EW}}\Big]  X^{(\overline B \ov
   K_1,\pi)}\Bigg\},
 \en
for $\ov B\to \ov K_1\pi$, and
 \be
  {\cal A}_{B^-\to K^-_1 K^0}
   &=&  \frac{G_F}{\sqrt{2}}\sum_{p=u,c}\lambda_p^{(d)}
    \Big[ \delta_{pu}\beta_2+ \alpha_4^p - \frac{1}{2}\alpha_{4,{\rm
    EW}}^p +\beta_3^p +\beta^p_{3,{\rm EW}} \Big]   X^{(\overline B \ov K_1,K)},  \non \\
   {\cal A}_{B^-\to K^0_1 K^-}
   &=&  \frac{G_F}{\sqrt{2}}\sum_{p=u,c}\lambda_p^{(d)}
    \Big[ \delta_{pu}\beta_2+ \alpha_4^p - \frac{1}{2}\alpha_{4,{\rm
    EW}}^p +\beta_3^p +\beta^p_{3,{\rm EW}} \Big]   X^{(\overline B \ov K,K_1)},  \non \\
    {\cal A}_{\ov B^0\to K^-_1 K^+}
   &=&  \frac{G_F}{\sqrt{2}}\sum_{p=u,c}\lambda_p^{(d)}\Bigg\{
    \Big[ \delta_{pu}\beta_1+\beta_4^p +\beta^p_{4,{\rm EW}} \Big]   X^{(\overline B K_1,K)}
    +f_Bf_{K_1}f_K\left[b_4^p-{1\over 2}b^p_{4,{\rm EW}}\right]_{KK_1}\Bigg\},  \non \\
     {\cal A}_{\ov B^0\to K^+_1 K^-}
   &=&  \frac{G_F}{\sqrt{2}}\sum_{p=u,c}\lambda_p^{(d)}\Bigg\{
    \Big[ \delta_{pu}\beta_1+\beta_4^p +\beta^p_{4,{\rm EW}} \Big]   X^{(\overline BK, K_1)}
   +f_Bf_{K_1}f_K\left[b_4^p-{1\over 2}b^p_{4,{\rm EW}}\right]_{K_1K}\Bigg\},  \non \\
     {\cal A}_{\ov B^0\to \ov K^0_1 K^0}
   &=&  \frac{G_F}{\sqrt{2}}\sum_{p=u,c}\lambda_p^{(d)}\Bigg\{
    \Big[ \alpha_4^p-{1\over 2}\alpha^p_{4,{\rm EW}}+\beta^p_3
    +\beta_4^p -{1\over 2}\beta^p_{3,{\rm EW}} -{1\over 2}\beta^p_{4,{\rm EW}}
    \Big]   X^{(\overline B \ov K_1,K)},  \non \\
    &+&   f_Bf_{K_1}f_K\left[b_4^p-{1\over 2}b^p_{4,{\rm EW}}\right]_{K\ov K_1}\Bigg\},  \non \\
     {\cal A}_{\ov B^0\to K^0_1 \ov K^0}
   &=&  \frac{G_F}{\sqrt{2}}\sum_{p=u,c}\lambda_p^{(d)}\Bigg\{
    \Big[ \alpha_4^p-{1\over 2}\alpha^p_{4,{\rm EW}}+\beta^p_3
    +\beta_4^p -{1\over 2}\beta^p_{3,{\rm EW}} -{1\over 2}\beta^p_{4,{\rm EW}}
    \Big]   X^{(\overline B \bar K,K_1)}, \non \\
    &+&   f_Bf_{K_1}f_K\left[b_4^p-{1\over 2}b^p_{4,{\rm EW}}\right]_{K_1\ov K}\Bigg\},
 \en
for $\ov B\to \ov K_1K$ and $\ov B\to K_1\ov K$, where
$\lambda_p^{(d)}\equiv V_{pb}V_{pd}^*$, $\lambda_p^{(s)}\equiv
V_{pb}V_{ps}^*$, and the factorizable amplitudes $X^{(\bar BA,P)}$
and $X^{(\bar B P,A)}$ are defined in Eq. (\ref{eq:XAP}). The decay
amplitudes for $\ov B\to b_1\pi$ and $b_1 \ov K$ are obtained from $\ov
B\to a_1\pi$ and $a_1 \ov K$ respectively by replacing $a_1\to b_1$.
Likewise, the expressions for $\ov B\to h_1\pi,h_1 \ov K$ decay
amplitudes are obtained by setting $(f_1\pi\to h_1\pi)$ and
$(f_1 \ov K\to h_1 \ov K)$.

The coefficients of the flavor operators $\alpha_i^p$  read
 \be \label{eq:alpha}
 \alpha_1(M_1,M_2) &=& a_1(M_1,M_2),\qquad \alpha_2(M_1M_2)=a_2(M_1M_2), \non \\
 \alpha_3^p(M_1M_2) &=& a_3^p(M_1M_2)-a_5^p(M_1M_2), \non \\
 \alpha_{\rm 3,EW}^p(M_1M_2) &=& a_9^p(M_1M_2)-a_7^p(M_1M_2), \non \\
 \alpha_4^p(M_1M_2) &=& \cases{ a_4^p(M_1M_2)+r_\chi^P
 a_6^p(M_1M_2); & for~$M_1M_2=AP$, \cr a_4^p(M_1M_2)-r_\chi^A
 a_6^p(M_1M_2); & for~$M_1M_2=PA$, } \non \\
 \alpha_{\rm 4,EW}^p(M_1M_2) &=& \cases{ a_{10}^p(M_1M_2)+r_\chi^P
 a_8^p(M_1M_2); & for~$M_1M_2=AP$, \cr a_{10}^p(M_1M_2)-r_\chi^A
 a_8^p(M_1M_2); & for~$M_1M_2=PA$, }
 \en
where $r_\chi^P$ and $r_\chi^A$ are defined before in Eq.
(\ref{eq:rchi}).


\end{document}